\documentclass[oldversion]{aa}
\usepackage{graphicx}
\usepackage{amssymb}
\usepackage[varg]{txfonts}

\begin{document}

   \title{Molecular gas chemistry in AGN}

   \subtitle{II. High-resolution imaging of SiO emission in NGC~1068: shocks or XDR? \thanks{Based on observations carried out with the IRAM Plateau de Bure Interferometer. 
IRAM is supported by INSU/CNRS (France), MPG (Germany) and IGN (Spain)}}

   \author{S.~Garc\'{\i}a-Burillo\inst{1}
			\and
	   A.~Usero\inst{1}
			\and
	   A.~Fuente\inst{1}
			\and
	  J.~Mart\'{\i}n-Pintado\inst{2}   
			\and 			
	  F.~Boone\inst{3}
	   		 \and 		
	   S.~Aalto\inst{4}
			\and
	   M.~Krips\inst{5}
			\and
	   R.~Neri\inst{5}
			\and
	   E.~Schinnerer\inst{6}
		  \and
	   L.~J.~Tacconi\inst{7}}

   \institute{
                   Observatorio Astron\'omico Nacional (OAN)-Observatorio de Madrid, Alfonso XII, 3, 28014-Madrid, Spain \\
			  \email{s.gburillo@oan.es, a.usero@oan.es, a.fuente@oan.es}
	 \and
	Centro de Astrobiolog\'{\i}a (CSIC-INTA), Ctra de Torrej\'on a Ajalvir, km 4, 28850 Torrej\'on de Ardoz, Madrid, Spain \\
			 \email{jmartin.pintado@iem.cfmac.csic.es}	
	\and
		Observatoire de Paris, LERMA, 61 Av. de l'Observatoire, 75014-Paris, France \\
			 \email{frederic.boone@obspm.fr}				 
	\and
		Department of Radio and Space Science with Onsala Observatory, Chalmers University of Technology, 439 94-Onsala, Sweden \\
			\email{saalto@chalmers.se}
	\and
		Institut de Radio Astronomie Millim\'etrique (IRAM), 300 rue de la Piscine, Domaine Universitaire de Grenoble, 38406-St.Martin d'H\`eres, France \\
			\email{neri@iram.fr}
	\and
	 Max-Planck-Institut f\"ur Astronomie, K\"onigstuhl, 17, 69117-Heidelberg, Germany \\
			 \email{schinner@mpia.de}
	\and
	 Max-Planck-Institut f\"ur extraterrestrische Physik, Postfach 1312, 85741-Garching, Germany \\
			 \email{linda@mpe.mpg.de}
}
	\date{Received ---; accepted ----}

\abstract{
{\it Context:}~This paper is part of a multi-species survey of line emission from the molecular gas in the
  circum-nuclear disk (CND) of the Seyfert 2 galaxy NGC~1068. 
Unlike in other active galaxies, the intensely star-forming regions in NGC~1068 and the CND can be resolved
with current instrumentation. This makes this galaxy an optimal test-bed to probe the effects of AGN on the molecular medium at $\sim100$~pc scales.    

{\it Aims:}~Single-dish observations have provided evidence that the abundance of silicon monoxide
  (SiO) in the CND of NGC~1068 is enhanced by 3--4 orders of magnitude with respect to the values typically measured in 
  quiescent molecular gas in the Galaxy. 
We aim at  unveiling the mechanism(s) underlying
  the SiO enhancement.

{\it Methods:}~We have imaged with the IRAM Plateau de Bure interferometer (PdBI) the emission of the SiO(2--1) (86.8~GHz)
and CN(2--1) (226.8~GHz) lines in NGC~1068 at $\sim150$~pc and $60$~pc spatial resolution, respectively. We
have also obtained complementary IRAM 30m observations of HNCO and methanol (CH$_3$OH) lines. These species
  are known as tracers of shocks in the Galaxy. 

{\it Results:}~SiO is detected in a disk of $\sim$400pc size around the AGN. SiO abundances in the CND of $\sim$(1--5)$\times$10$^{-9}$ are about 1--2 orders of magnitude above those measured in the starburst ring. 
The overall abundance of CN in the CND is high: $\sim$(0.2--1)$\times$10$^{-7}$.  The abundances of SiO and CN are enhanced at the extreme velocities of gas associated with non-circular motions close to the AGN ($r<$70~pc). 
 On average, HNCO/SiO and CH$_3$OH/SiO line ratios in the CND are similar to those measured in prototypical shocked regions in our Galaxy. However, the strength and abundance of CN in NGC~1068 can be explained neither by shocks nor by PDR chemistry. Abundances measured for CN and SiO, and the correlation of CN/CO and SiO/CO ratios with hard X-ray irradiation, suggest that the CND of NGC~1068 has become a giant X-ray dominated region (XDR).

{\it Conclusions:}~The extreme properties of molecular gas in the circum-nuclear molecular disk of NGC~1068  result from the interplay between different processes directly linked to
nuclear activity.  Results presented in this paper highlight, in particular, the footprint of shocks and X-ray irradiation on the properties of molecular gas in this Seyfert. Whereas XDR chemistry offers a simple explanation for CN and SiO in NGC~1068, the relevance of shocks deserves further scrutiny. The inclusion of dust grain chemistry would help solve the controversy regarding the abundances of other molecular species, like HCN, which are under-predicted by XDR models.
 }


   \maketitle

\section{Introduction}\label{Introduction}

Nuclear activity and intense star formation can shape the excitation and chemistry of molecular gas in the circum-nuclear disks (CND) of galaxies.  The effect on molecular gas properties of strong radiation fields (UV and X-rays) and the injection of mechanical energy (by gas outflows or jet-ISM interactions) are key ingredients in the feedback of activity. Molecular line observations can be used to study the nature of the dominant source of energy in galaxies hosting both active galactic nuclei (AGN) and intense star formation . Furthermore, nearby galaxies can serve as local templates of distant galaxies where these phenomena can be deeply embedded.

Some of the tracers of the dense molecular gas phase, which are commonly used in extragalactic research, can be heavily affected by the feedback of AGN activity. In particular the lines of HCN (the most widely observed tracer of the dense molecular medium) appear as {\it over-luminous} in the CND of many Seyfert galaxies with respect to other tracers of the dense molecular gas (Tacconi et al.~\cite{Tac94}; Kohno et al.~\cite{Koh01}; Usero et al.~\cite{Use04}, hereafter U04; Krips et al.~\cite{Kri08}). Doubts have been thus cast on the universality of the conversion factor between the luminosity of the HCN(1--0) line, $L_{\rm HCN(1-0)}$, and the mass of dense molecular gas in active galaxies (Graci\'a-Carpio et al.~\cite{Gra06, Gra08}; Garc\'{\i}a-Burillo et al.~\cite{Gar06}; Krips et al.~\cite{Kri08}). X-rays can efficiently process large column densities of molecular gas around AGNs, producing X-ray dominated regions (XDR). It has been proposed that the abundances of certain ions, radicals and molecular species, like HCN, can be enhanced in XDR (Lepp \& Dalgarno~\cite{Lep96}; Maloney et al.~\cite{Mal96}; Meijerink \& Spaans~\cite{Meij05}; Meijerink et al.~\cite{Meij07}). However, it is still controversial whether pure gas-phase XDR models are able to enhance HCN abundances in AGN to the level imposed by observations.

The inclusion of dust grain chemistry, not taken into account by gas-phase XDR schemes, could solve the HCN controversy in AGN by enhancing the abundance of this molecule. Besides HCN, other molecular species can also undergo significant changes in their abundances due to dust grain processing. X-rays can evaporate small ($\sim$10~\AA) silicate grains (Voit~\cite{Voi91}). This can increase the Si fraction in the gas phase and then considerably enhance the abundance of SiO in X-ray irradiated molecular gas (Mart\'{\i}n-Pintado et al.~\cite{Mar00}; U04; Garc{\'{\i}}a-Burillo et al.~\cite{Gar08}; Amo-Baladr\'on et al.~\cite{Amo09}). Furthermore, mechanical sputtering of dust grains in molecular shocks is an additional source of dust grain chemistry. Large SiO abundances have been found in the nuclei of a number of non-AGN galaxies where interferometer maps have been key to identifying large-scale molecular shocks (Garc\'{\i}a-Burillo et al.~\cite{Gar00,Gar01}; Usero et al.~\cite{Use06}). 

To address the role of X-rays and shocks in shaping the chemistry of molecular gas in AGN requires a multi-species approach. Unlike SiO, other tracers of the dense molecular gas, like CN, are not expected to be highly abundant in shocks (Mitchell~\cite{Mit84}; Fuente et al.~\cite{Fue05}; Rodr\'{\i}guez-Fern\' andez et al.~\cite{Rod10}). However, there is a consensus supported by observations and theoretical models that the CN radical is a privileged tracer of highly ionized molecular gas, typical of photon or X-ray dominated regions (PDR or XDR) (Boger \& Sternberg~\cite{Bog05}; Fuente et al.~\cite{Fue93, Fue08}; Janssen et al.~\cite{Jan95}; Lepp \& Dalgarno~\cite{Lep96}; Meijerink \& Spaans~\cite{Meij05}; Meijerink et al.~\cite{Meij07}).  Observations 
of tracers like SiO and CN can then be used to quantify the relevance of shocks and XDR chemistry in AGN.  Furthermore the high-spatial resolution provided by interferometers is paramount to spatially discriminate between the different chemical environments (star formation vs AGN activity) that can coexist in a single galaxy. 


NGC\,1068 is the strongest nearby Seyfert 2 galaxy, and as such is a prime candidate for studying the feeding and the feedback of activity using molecular line observations. Tacconi et al.~(\cite{Tac94}) and Schinnerer et al.~(\cite{Sch00}; hereafter S00) used the Plateau de Bure Interferometer (PdBI) to map at high-resolution (1--4$\arcsec$) the emission of molecular gas in the central $r$\,$\sim$1.5-2~kpc disk using HCN and CO lines. The CO maps spatially resolve the distribution of molecular gas in the disk, showing a prominent starburst ring (hereafter SB ring) of $\sim$1-1.5\,kpc--radius, which contributes significantly to the total CO luminosity of NGC\,1068 (see also Planesas et al.~\cite{Pla91} and Baker~\cite{Bak00}, hereafter B00). Furthermore, CO emission is detected in a central $r$\,$\sim$200\,pc CND surrounding the AGN. In contrast to CO, emission from the HCN(1--0) line, which is also spatially resolved in the Tacconi et al.'s maps, arises mainly
from the CND. Tacconi et al.~(\cite{Tac94}) derived a high HCN/CO intensity ratio ($\sim$1) in the CND. This is about a factor of 5--10 higher than the average ratio measured in the SB ring (Usero et al.~in prep). 
Radiative transfer calculations based on multi-line observations of HCN and CO showed that the abundance of HCN relative to CO was significantly enhanced in the CND: HCN/CO$\sim$10$^{-3}$ (Sternberg et al.~\cite{Ste94}).

U04 used the IRAM 30m telescope to observe with low to moderate spatial resolutions ($\sim$10--30$\arcsec$) the emission of eight molecular species in the CND of NGC\,1068. The global analysis of the survey, which includes several lines of SiO, CN, HCO, H$^{13}$CO$^{+}$, H$^{12}$CO$^{+}$, HOC$^{+}$, HCN, CS, and CO, indicated that the bulk of the molecular gas emission in the CND of NGC\,1068 could be interpreted as coming from a giant XDR. More recently, P\'erez-Beaupuits et al.~(\cite{Per07, Per09}) analyzed new single-dish data obtained for several rotational lines of HCO$^+$, HNC, CN and HCN, and came to similar conclusions as U04. However, the low spatial resolution of these single-dish observations are not optimal to precisely disentangle the contributions of the starburst and the CND in NGC~1068.

In this paper, we use the high-spatial resolution ($\sim$1--3$\arcsec$) afforded by PdBI to map the emission of the ($v$=0, $J$=2--1) line of SiO and the  $N$=2--1 transition of CN in the central $r$\,$\sim$1.5-2~kpc disk of NGC~1068. The spatial resolution of the new observations, an order of magnitude higher than that of the 30m survey of U04, allow us to neatly separate the emission of the SB ring from that of the CND. Furthermore, the SiO and CN PdBI maps, together with complementary single-dish data obtained in CH$_ 3$OH and HNCO lines, are used to study the chemical differentiation inside the CND through an analysis of line ratio maps. Line ratios are interpreted with the help of one-phase large velocity gradient (LVG) models. While the emission of the different lines and species analyzed in this work likely comes from  regions characterized by different physical conditions, the adopted one-phase approach allows us to explore the existence of overall trends in the chemical abundances of these species, using a minimum set of free parameters in the fit. We explore the dependence of line ratios with the illumination of molecular gas by the X-ray AGN source. We also explore the link between shock chemistry and gas kinematics in the CND.

We describe in Sect.~\ref{Observations} the observations, including high-resolution SiO, CN and CO maps obtained with the PdBI, single-dish data obtained with the 30m telescope as well as X-ray images taken by Chandra of NGC\,1068. Sect.~\ref{Continuum} presents the continuum maps derived at 86.8\,GHz and 226.8\,GHz. The distribution and kinematics of molecular gas derived from SiO and CN are described in Sects.~\ref{Distribution} and ~\ref{Kinematics}.  SiO and CN abundances are discussed in Sect.~\ref{$X$}.  In Sect.~\ref{Models} we interpret the line ratio maps in terms of two different chemical scenarios (shocks and XDR). The main conclusions of this work are summarized in Sect.~\ref{Summary and conclusions}. We assume a distance to NGC\,1068 of $D\sim$14~Mpc (Bland-Hawthorn et al.\cite{Bla97}); this implies a spatial scale of $\sim$70~pc/$\arcsec$.

\section{Observations}\label{Observations}

\subsection{Interferometric maps}\label{PdbI}

Observations of NGC\,1068 were carried out with the PdBI array (Guilloteau et al.~\cite{Gui92}) between December 2004 and February 2005. We used the BC configurations and six antennas in dual frequency mode. We simultaneously observed the ($v$=0, $J$=2--1) line of SiO (at 86.847\,GHz) and the $N$=2--1 transition of CN. The CN(2--1) transition is split up into 18 hyperfine lines blended around three groups ($J$=5/2--3/2, $J$=3/2--1/2 and $J$=3/2--3/2). We have observed the two strongest groups of lines at 226.9\,GHz ($J$=5/2--3/2) and 226.7\,GHz ($J$=3/2--1/2), hereafter referred to as high frequency (HF) and low frequency (LF) CN lines, respectively. During the observations the spectral correlator was split in two halves centered at 86.800\,GHz and 226.767\,GHz. This choice allowed us to cover the SiO line at 3~mm and the CN lines at 1~mm. The $J$=1--0 line of H$^{13}$CO$^+$ (at 86.754\,GHz) is simultaneously covered by the 3~mm setup. Rest frequencies were corrected for the recession velocity initially assumed to be $v_{o}(HEL)$=1137\,km~s$^{-1}$.  The correlator configuration covers a bandwidth of 580\,MHz for each setup, using four 160\,MHz-wide units with an overlap of 20\,MHz; this is equivalent to 2000\,km~s$^{-1}$(770\,km~s$^{-1}$) at 86.8\,GHz (226.8\,GHz). Observations were conducted in single pointings of sizes 56$^{\prime\prime}$ and 21$^{\prime\prime}$ at 3~mm and 1~mm, respectively, centered at $\alpha_{2000}$=02$^{h}$42$^{m}$40.71$^{s}$ and $\delta_{2000}$=--00$^{\circ}$00$^{\prime}$47.94$\arcsec$. The latter corresponds to the nominal position of the AGN core, as determined from different VLA and VLBI radio continuum images of the galaxy (e.g., Gallimore et al.~\cite{Gal96}). Visibilities were obtained using on-source integration times of 20 minutes framed by short ($\sim$\,2\,min) phase and amplitude calibrations on the nearby quasars 0235+164 and 0336-019. The absolute flux scale in our maps was derived to a 10$\%$ accuracy based on the observations of primary calibrators whose fluxes were determined from a combined set of measurements obtained at the 30m telescope and the PdBI array. The bandpass calibration was carried out using NRAO150 and 1749+096 and is accurate to better than 5$\%$.

The image reconstruction was done using
standard IRAM/GILDAS software (Guilloteau \& Lucas~\cite{Gui00}). We used natural weighting and no taper to generate the SiO line map with a size of 64$\arcsec$ and 0.25$\arcsec$/pixel sampling; the corresponding synthesized beam is $3.5''\times 2.1''$, $PA$=29$^{\circ}$. We also used natural weighting to generate the CN map with a size of 32$\arcsec$ and 0.12$\arcsec$/pixel sampling; this enables us to achieve a spatial resolution of $\sim$1$''$ ($1.5''\times 0.8''$, $PA$=19$^{\circ}$). The conversion factors between Jy\,beam$^{-1}$ and K are 22\,K~Jy$^{-1}$~beam at 86.8\,GHz, and 21\,K~Jy$^{-1}$~beam at 226.8\,GHz. The point source sensitivities are derived from emission-free channels. They are 1.5\,mJy~beam$^{-1}$ in 2.5~MHz-wide channels at 3mm, and  7.6\,mJy~beam$^{-1}$ in 6.25~MHz-wide channels at 1mm. Images of the continuum emission of the galaxy have been obtained by averaging those channels free of line emission at both frequency ranges. The corresponding point source sensitivities for continuum images are 0.3\,mJy~beam$^{-1}$ at 86.8\,GHz and 0.7\,mJy~beam$^{-1}$ at 226.8\,GHz.

In this work we use the 1.8$\arcsec$ resolution CO(1--0) interferometer maps of NGC\,1068 obtained with the PdBI by S00. We list in Table~1 the relevant parameters for the SiO and CN PdBI observations presented in this paper.

\begin{table*}[!htp]
\begin{center}
\caption{Observational parameters of the SiO and CN PdBI data.}    
\label{Intensities}
{
{
\begin{tabular}{lcc}
\hline
\hline
\noalign{\smallskip}
	& SiO &  CN  \\ 
\noalign{\smallskip}
\hline
\noalign{\smallskip}
Line	& $v$=0, $J$=2--1  & $N$=2--1 ($J$=5/2--3/2 (HF), $J$=3/2--1/2 (LF))   \\
\noalign{\smallskip}
\hline
\noalign{\smallskip}

Frequency & 86.8~GHz & 226.9~GHz (HF), 226.7~GHz (LF) \\
\noalign{\smallskip}
\hline

\noalign{\smallskip}
Beam		&  $3.5''\times 2.1''$, $PA$=29$^{\circ}$ &  $1.5''\times 0.8''$, $PA$=19$^{\circ}$ \\
\noalign{\smallskip}
\hline
\noalign{\smallskip}

1-$\sigma$ (rms) & 1.5\,mJy~beam$^{-1}$ ($\Delta \nu$=2.5~MHz) & 7.6\,mJy~beam$^{-1}$ ($\Delta \nu$=6.25~MHz)  \\
\noalign{\smallskip}
\hline

\noalign{\smallskip}

Field of view & 56$^{\prime\prime}$  & 21$^{\prime\prime}$  \\
\noalign{\smallskip}
\hline
\noalign{\smallskip}

$T_{mb}$/$S$ & 22\,K~Jy$^{-1}$~beam & 21\,K~Jy$^{-1}$~beam \\
\noalign{\smallskip}
\hline

\end{tabular}
}

}
\end{center}
\label{table1}
\end{table*}



\begin{figure}
   \centering
   \includegraphics[width=7cm]{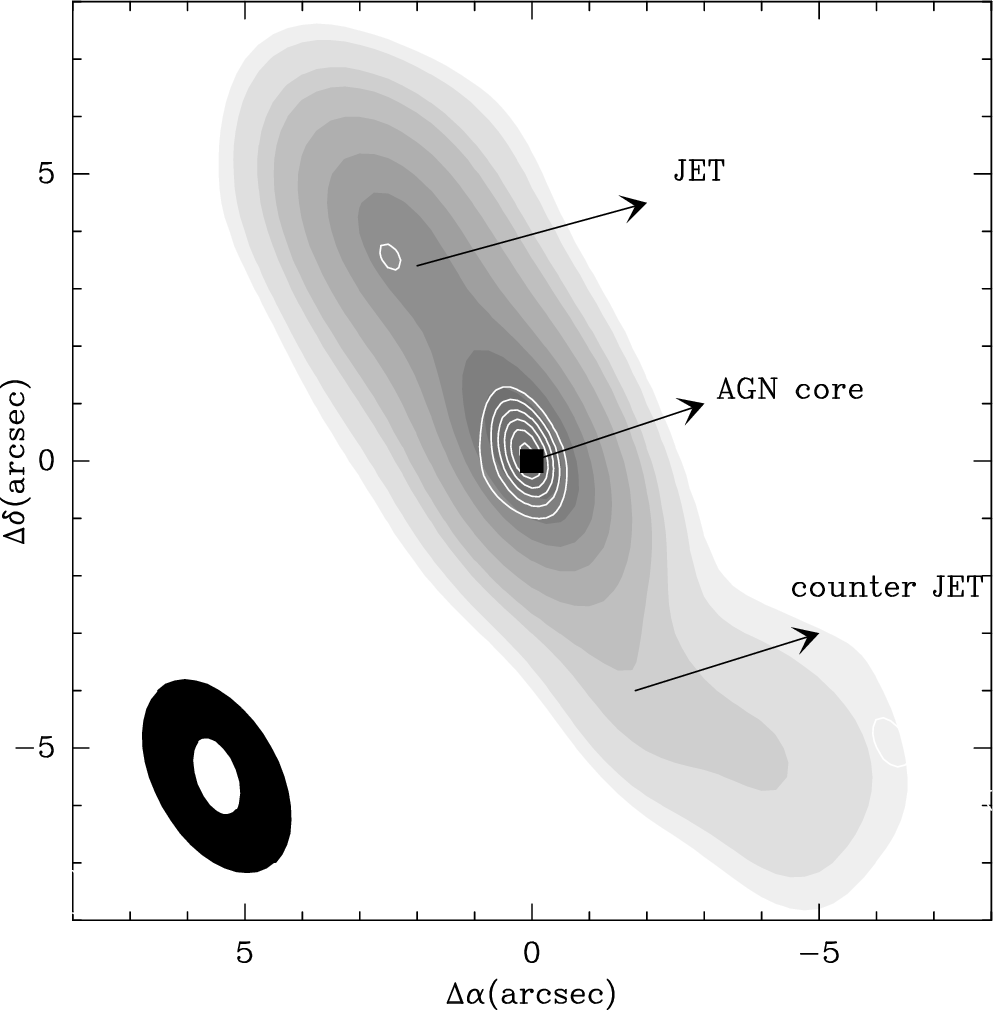}
   \caption{The continuum maps obtained with the PdBI towards the nucleus of NGC~1068 at 3~mm (grey scale contours: 3\%, 5\%, 10\%, 15\%, 25\% to 85\% in steps of 15\% of the maximum=39.8~mJy~beam$^{-1}$) and 1~mm (white contours: 20\% to 95\% in steps of 15\% of the maximum=14.2~mJy~beam$^{-1}$). The lowest contours correspond to $\sim$4$\sigma$ levels at both frequencies. ($\Delta\alpha$,~$\Delta\delta$)--offsets are relative to the phase tracking center ($\alpha_{2000}$=02$^{h}$42$^{m}$40.71$^{s}$, $\delta_{2000}$=--00$^{\circ}$00$^{\prime}$47.94$\arcsec$), which coincides with the position of the AGN (filled square). Black-filled and white-filled ellipses show the beams at 3~mm (3.5$\arcsec\times$2.1$\arcsec$ at $PA$=29$^{\circ}$) and 1~mm (1.5$\arcsec\times$0.8$\arcsec$ at $PA$=19$^{\circ}$), respectively.} 
              \label{continuum}
\end{figure}


\begin{figure}
   \centering
   \includegraphics[width=8cm]{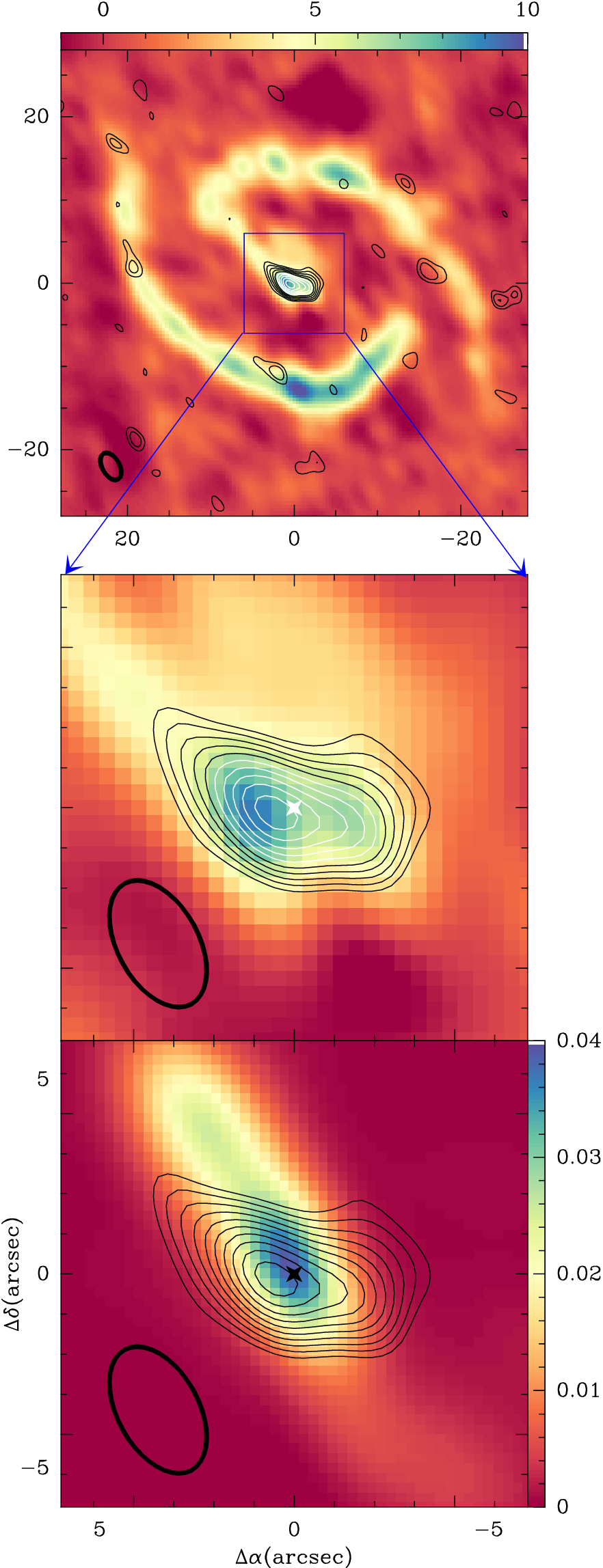}
    \caption{{\bf a)}~({\it Upper panel})~The SiO integrated intensity map (contour levels are 3$\sigma$ to 12$\sigma$ in steps of 1$\sigma$=0.082~Jy~km~s$^{-1}$~beam$^{-1}$) is overlaid on the CO(1--0) integrated intensity map of S00 (color scale as shown in units of Jy~km~s$^{-1}$~beam$^{-1}$). {\bf b)}~({\it Middle panel})~The same as {\bf a)} but showing a
zoomed view on the inner 12$\arcsec$ around the AGN (identified by the cross). {\bf c)}~({\it Lower panel})~The SiO map contours are overlaid on the 3~mm continuum map (color scale) of Fig.~\ref{continuum}. The SiO beam is shown in all the panels by an ellipse.}
              \label{SiO-maps}
\end{figure}


\begin{figure}
   \centering
   \includegraphics[width=7cm]{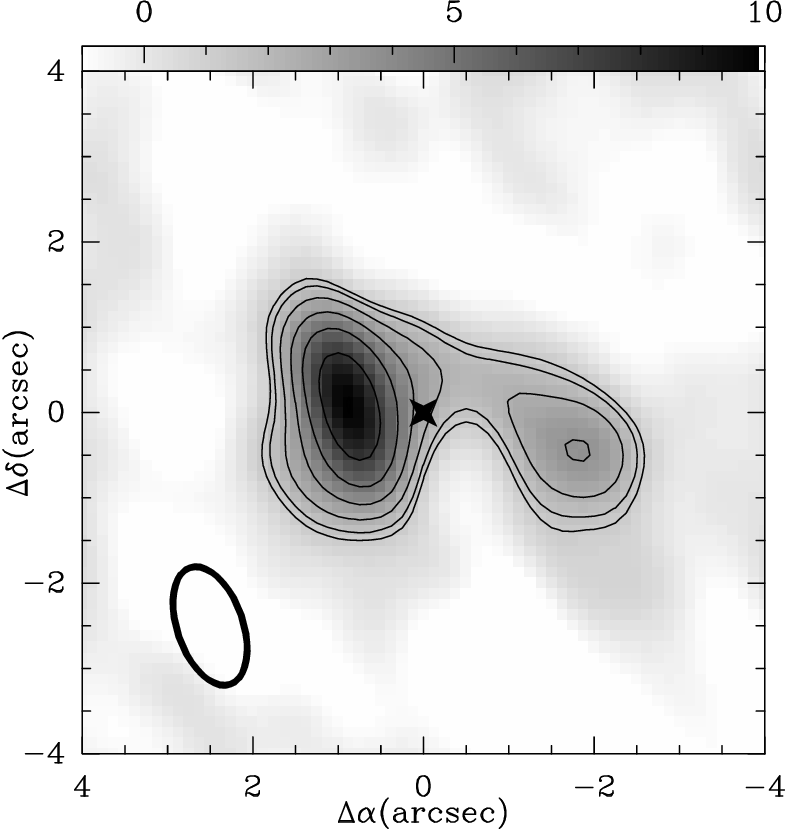}
    \caption{~The CN integrated intensity map. Contour levels are 3$\sigma$, 4$\sigma$, 6$\sigma$, 
 9$\sigma$, 13$\sigma$ to 25$\sigma$ in steps of 6$\sigma$, with 1$\sigma$=0.40~Jy~km~s$^{-1}$~beam$^{-1}$. The AGN is identified by the cross and the CN beam is shown by an ellipse.}
 \label{CN-maps}
\end{figure}

\subsection{Single-dish spectra}\label{30m}

New CH$_3$OH and HNCO observations of the nucleus of NGC\,1068 were carried out in July 2006 with the IRAM 30m telescope at Pico de Veleta (Spain). We observed the 2$_k$--1$_k$ group of transitions of CH$_3$OH. This group is a blended set of four lines. Hereafter we consider the frequency of the  2$_{0}$-1$_0$A$^+$ line (96.741\,GHz) as velocity reference.
We observed the higher frequency group of transitions of methanol at 2~mm (denoted as 3$_k$--2$_k$), a blended set of eight lines.  The frequency of the 3$_{0}$-2$_0$A$^+$ line (145.103\,GHz) is taken as  velocity reference. We also observed the 4$_{04}$--3$_{03}$ rotational transition of HNCO (hereafter designated as 4--3) at 87.925\,GHz. The corresponding beam sizes of the telescope are $\sim$28$\arcsec$ at 3~mm and 17$\arcsec$ at 2~mm. The 3\,mm and 2\,mm SIS receivers of the 30m telescope were tuned to the redshifted frequencies of the lines around $v_{o}(HEL)$=1137\,km~s$^{-1}$. The velocity range covered was $\rm 1600\,km\,s^{-1}$ for the 3\,mm lines and $\rm 900\,km\,s^{-1}$ for the 2\,mm lines. The wobbler switching mode was used to obtain flat baselines with a maximum beam throw of 4$\arcmin$.

Typical system temperatures during the observations were $\sim$200\,K at 3\,mm and $\sim$300\,K at 2\,mm. All receivers were used in single side-band mode (SSB), with a high rejection of the image band.
The calibration accuracy is better than 20$\%$. Pointing of the 30m telescope was regularly checked every 1.5 hours by observing a nearby continuum source; we found an average rms pointing error of 2--3$\arcsec$ during the observations. 

Throughout the paper, line intensities are given in antenna temperature scale, $T_{\rm a}^{*}$. The $T_{\rm a}^{*}$ scale relates to the main beam temperature scale, $T_{\rm mb}$, by the equation $T_{\rm mb} = (F_{\rm eff}/B_{\rm eff}) T_{\rm a}^{*}$, where $F_{\rm eff}$ and $B_{\rm eff}$ are, respectively, the forward and beam efficiencies of the telescope at a given frequency. For the IRAM 30m telescope $F_{\rm eff}/B_{\rm eff} = 1.26$ (1.48) at 86\,GHz (145\,GHz).

\section{Continuum maps}\label{Continuum}

Figure~\ref{continuum} shows the continuum maps derived at 86.8\,GHz and 226.8\,GHz in the nucleus of 
NGC~1068. The morphology of the maps is similar to that described by Krips et al.~(\cite{Kri06}), who observed the continuum emission at the nearby frequencies 110\,GHz and 230.5\,GHz. The emission consists of a central component ({\it AGN core}), a NE elongation ({\it jet}), and a SW elongation ({\it counter-jet}). The {\it counter-jet} and the {\it jet} are only detected at 3~mm. 

We have used the GILDAS task UV-FIT to fit
the continuum visibilities at both frequencies using a set of three point sources. Most of the flux comes from the compact source located at the position of the VLBI radio core at ($\Delta\alpha$,~$\Delta\delta$)$\sim$(0$\arcsec$, 0$\arcsec$). The {\it AGN core} has a flux of 39.8$\pm$0.3~mJy and 16.0$\pm$0.7~mJy at 86.8\,GHz and 226.8\,GHz, respectively. 
The NE {\it jet} component is fitted by a point source of 23.8$\pm$0.3~mJy at 86.8\,GHz, located at ($\Delta\alpha$,~$\Delta\delta$)=(2.6$\arcsec$, 3.4$\arcsec$). 
The 3~mm {\it counter-jet} point source at ($\Delta\alpha$,~$\Delta\delta$)=(--1.6$\arcsec$, --4.0$\arcsec$) has a flux of 4.7$\pm$0.3~mJy. There are nevertheless hints of extended emission, mostly at 3~mm, not fully accounted for by the point source fitting (Krips et al.~\cite{Kri06}). Taking into account the differences in frequency, spatial resolution and fitting functions used, the solution described above is in agreement with that found by Krips et al.~(\cite{Kri06}) (see their Table~1 for a detailed description).

\section{Molecular gas distribution}\label{Distribution}

\subsection{SiO distribution}\label{SiO-dist}

Figure~\ref{SiO-maps} shows the SiO(2--1) intensity map obtained by integrating the emission in velocity channels from $v$\,=\,967 to 1307~km~s$^{-1}$. In this velocity range line emission is detected at significant $\geq$3$\sigma$ levels on scales larger than the beam inside the PdBI field of view. We overlay the SiO(2--1) map on the CO(1--0) map of S00. The CO map fairly represents the overall molecular gas distribution in NGC\,1068. S00 estimated that the CO PdBI map recovers $\simeq$80$\%$ of the total flux, based on the comparison between the CO flux measured by the PdBI and the flux derived from the CO maps obtained through the combination of the BIMA array and  NRAO single-dish data of Helfer \& Blitz~(\cite{Hel95}). For the sake of comparison, the spatial resolution of the CO(1--0) map shown in Fig.~\ref{SiO-maps}, originally $\sim$1.8$\arcsec$, has been degraded to that of the SiO data. In Fig.~\ref{SiO-maps} we also compare the SiO map to the morphology of the jet derived from the 3~mm continuum data.

Unlike CO, most of the total SiO emission detected inside the PdBI primary beam comes from a circum-nuclear molecular disk located around the AGN, referred to as CND. The integrated flux of SiO in the CND is $\sim$2~Jy~kms$^{-1}$. This is $\sim$74$\%$ of the total flux detected by U04 at 28$\arcsec$ spatial resolution with the 30m telescope towards the AGN. U04 estimate that $\simeq$75$\%$ of the SiO emission of their AGN spectrum comes from the CND itself, with a residual $\simeq$25$\%$ contribution from the SB ring. Taken together, these
estimates indicate that the PdBI recovers practically all of the SiO flux in the CND. Little SiO emission is detected from the SB ring in the PdBI map. A few SiO clumps, showing sizes $\sim$beam and intensities at the 3--4$\sigma$ level, are distributed along the SB ring.

The overall distribution of SiO in the CND, spatially resolved by the PdBI beam, shows an east-west elongation. The deconvolved full size of the emission at a 4$\sigma$ level is $\sim$400~pc along the east-west axis. The CND is marginally resolved along the north-south axis; the corresponding deconvolved size is $\leq$150~pc. These sizes are comparable with those derived from the higher resolution 1--0 and 2--1 CO maps of the CND published by S00, as well as with those derived from the CN(2--1) map described in Sect.~\ref{CN-dist}. The SiO emission is slightly asymmetric around the AGN: the SiO emission peak is at ($\Delta\alpha$,~$\Delta\delta$)$\sim$(0.5$\arcsec$, 0$\arcsec$). Instead, the CO map shows a maximum at ($\Delta\alpha$,~$\Delta\delta$)$\sim$(1.2$\arcsec$, 0$\arcsec$) (see Fig.~1 of S00). The latter corresponds to the E CO knot, in the notation of S00. The E CO knot is thus shifted  $\sim$50~pc east relative to the E SiO knot. 


\subsection{CN distribution}\label{CN-dist}

Both lines of CN (LF and HF, as defined in Sect.~\ref{Observations}) are detected in the CND of NGC\,1068. As expected, none of the CN lines are detected towards the SB ring, as this region is located well beyond the edge of the primary beam of the PdBI at 226.8~GHz. The HF/LF intensity ratio shows no significant spatial variation inside the CND. The average ratio, $\sim$1.7, is in close agreement with the value expected in the limit of optically thin emission (HF/LF$\sim1.64\pm0.14$). In the remaining of this paper, we will use the higher S/N ratio maps of the HF line.

Figure~\ref{CN-maps} shows the intensity map of the HF line of CN(2--1) obtained towards the CND of NGC~1068. We have integrated the emission in velocity channels from $v$\,=\,955 to 1277~km~s$^{-1}$, i.e., similar to the velocity range used to derive the SiO map. With this choice we encompass all significant line emission and at the same time avoid blending with the LF CN(2--1) line. At this spatial resolution ($\sim$1$\arcsec$), the overall morphology of the CND traced by CN roughly resembles that seen in CO lines. At close sight there are noticeable differences between these tracers, however. The two CN knots, connected by a bridge of emission on the northern side of the CND, lie at ($\Delta\alpha$,~$\Delta\delta$)$\sim$(0.9$\arcsec$, 0$\arcsec$) and ($\Delta\alpha$,~$\Delta\delta$)$\sim$(--1.7$\arcsec$, --0.5$\arcsec$).
As for SiO, the CN knot located east is significantly closer to the AGN compared to the corresponding CO knot.

\section{Molecular gas kinematics}\label{Kinematics}

\subsection{Background from previous work}\label{Previous}

The overall kinematics of molecular gas in the disk of NGC\,1068 were modeled by S00 and B00. One of the scenarios advanced by S00 invokes the presence of two embedded bars: an outer oval and a nuclear bar of $\sim$17~kpc and  2.5~kpc deprojected diameters, respectively. The two bars are coupled through resonance overlapping, so that the corotation of the nuclear bar coincides  with the outer inner Lindblad resonance (ILR) of the oval. The gas response inside $r\sim$5$\arcsec$(350~pc), a region that encompasses the whole CND detected in SiO and CN, would lie between the ILRs of the nuclear bar. 

S00 and B00 proposed also a nuclear warp model to alternatively account for the 
anomalous kinematics of molecular gas in the CND. The non-coplanar gas response in the nuclear region could have been caused by the interaction of the molecular disk with the jet or with the associated ionization cone. B00's model proposes a warped molecular disk that evolves from a low inclination ($i\sim$30$^{\circ}$) at $r\sim$2.5$\arcsec$  to a highly-inclined disk ($i\sim$77$^{\circ}$) at $r\sim$0.5$\arcsec$. This configuration naturally explains the steep velocity gradient ({\it high velocities}) measured in the major axis position-velocity (p-v) diagram of CO at small radii ($r\leq$2$\arcsec$).
 In particular, S00 favor a hybrid model where the gas response to the bar is combined with a nuclear warp. Besides being able to reasonably fit the gas kinematics, the main advantage of the hybrid solution over the pure bar model is that the first explains the asymmetric excitation of molecular gas evidenced by the remarkably different $R$=2--1/1--0 line ratios measured in the east and the west CO lobes. B00 argues that the east lobe, the one showing the higher R, corresponds to the directly X-ray illuminated surface of a warped disk, whereas the west lobe corresponds to the back of the warped disk, characterized by a lower R. The illumination of the CND gas by X-rays may also explain the particular chemistry of molecular clouds analyzed in Sect.~\ref{SiO-CN-CO-CND}.
   
 The kinematics of molecular gas in the inner $r\sim$50~pc region are also complex, as recently revealed by the 2.12$\mu$m H$_2$ 1--0 S(1) map of M\"uller-S{\'a}nchez et al.~(\cite{Mue09}). These data, mostly sensitive to hot (T$_K\simeq$10$^3$~K) and moderately dense (n(H$_2$)$\simeq$10$^3$~cm$^{-3}$) molecular gas, show the existence of a structure bridging  the CND and the central engine. This connection is made through highly elliptical streamers detected in H$_2$ lines. Davies et al.~(\cite{Dav08}) used these data to study the large-scale kinematics of molecular gas in the CND itself and concluded that the whole CND is a lopsided ring in expansion (Krips et al., in prep.). The scenario of an expanding ring, first advanced by Galliano \& Alloin~(\cite{Gal02}), could be linked to the onset of a nuclear warp instability.

\subsection{SiO kinematics}\label{Kinematics-SiO}

\begin{figure}[t!]
   \centering
   \includegraphics[width=7cm]{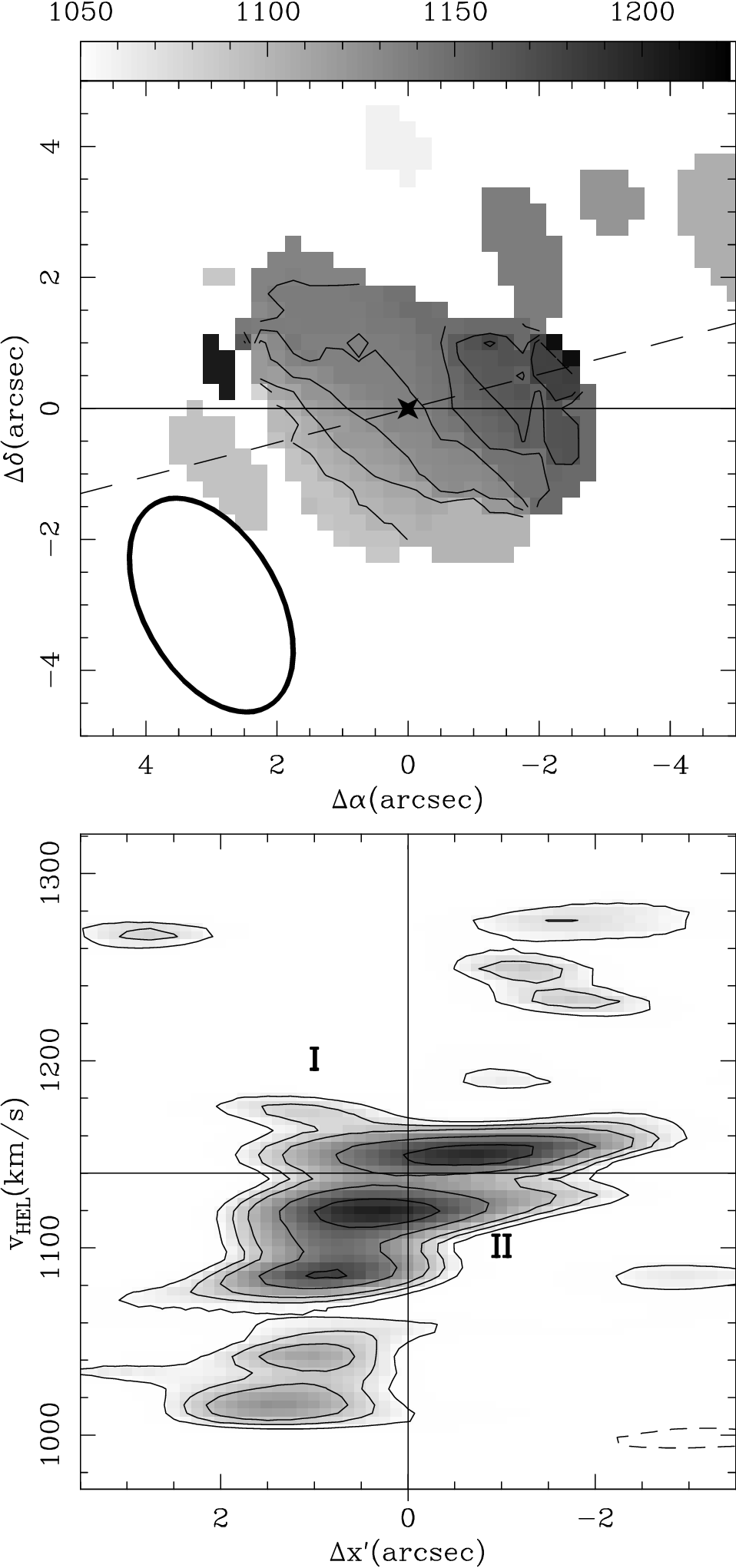}
   \caption{{\bf a)}~({\it Upper panel}) SiO isovelocities contoured over false-color velocity maps. Velocities span the range [1050~km~s$^{-1}$, 1225~km~s$^{-1}$] in steps of 25~km~s$^{-1}$. Velocity scale is HEL. The AGN position is marked with a star. The position of the kinematic major axis derived from SiO at $PA\sim$105$^{\circ}$ is indicated by the dashed line. An ellipse shows the beam. {\bf b)}~({\it Lower panel}) Position-velocity diagram of SiO along 
 $PA$=90$^{\circ}$. Contour levels are -3, 3, 3.75, 4.5 to 9 in steps of 1.5~mJy~beam$^{-1}$. The position of the AGN ($\Delta$x'=0$\arcsec$) and $v_{sys}$(HEL)=1140~km~s$^{-1}$ are highlighted. We indicate the position of quadrants I and II discussed in the text.}
              \label{SiO-velocities}
\end{figure}

Figure~\ref{SiO-velocities}{\it a} shows the isovelocity contour map of the CND of NGC\,1068 derived from SiO. Isovelocities have been obtained using a 3$\sigma$ clipping on the data cube. The velocity centroid of SiO emission towards the AGN is $v_{AGN}\sim$1122$\pm$10~km~s$^{-1}$ in HEL scale.  Gas velocities are red ($v>v_{AGN}$) on the northwest side of the CND, while blue velocities ($v<v_{AGN}$) appear southeast. This picture is consistent with a spatially resolved rotating molecular structure. The kinematic major axis is fitted to $PA\sim$105$\pm$10$^{\circ}$ to expand the maximum range of SiO radial velocities at the edge of the CND ($r\simeq$2.5$\arcsec$). This solution lies within the wide range of $PA$ values derived from the photometric fitting of the disk and, also, from the observed stellar and gas kinematics ($PA\sim$80$^{\circ}$--125$^{\circ}$; Brinks et al.~\cite{Bri97}; Dehnen et al.~\cite{Deh97}; S00; B00; Emsellem et al.~\cite{Ems06}; Gerssen et al.~\cite{Ger06}). We note that SiO isovelocities are twisted to larger position angles ($PA\sim$135$^{\circ}$) at smaller radii ($r\leq$1$\arcsec$), which already suggests the presence of non-circular and/or non-coplanar motions in the CND. A similar isovelocity twist is seen in other molecular tracers, as pointed out by B00. Outside the CND, the large-scale stellar and gas kinematics indicate that the major axis of the disk is oriented east-west (Emsellem et al.~\cite{Ems06}). We therefore adopt in the following PA=90$^{\circ}$ as the best guess for the major axis orientation. 

Figure~\ref{SiO-velocities}{\it b} shows the SiO p-v diagram along the major axis. The p-v diagram is mostly symmetric for the lowest contour levels around the position of the AGN ($\Delta$x'=0$\arcsec$), and with respect to a radial velocity $v$=$v_{sys}$(HEL)=1140$\pm$10~km~s$^{-1}$. We therefore adopt the first as the dynamical center and the second as the systemic velocity $v_{sys}$ of the galaxy. This value of $v_{sys}$ is in good agreement with all previous determinations of the galaxy's receding velocity ($\sim$1137$\pm$3~km~s$^{-1}$; e.g., de Vaucouleurs et al.~\cite{deV91}; Huchra et al.~\cite{Huc99}). The highest contour levels of the p-v plot are shifted east, however; this reflects the overall east-west asymmetry of the CND described in Sect.\ref{SiO-dist}.

The steep velocity gradient ({\it high velocities}) measured along the major axis at $r\leq$2$\arcsec$ 
can be interpreted at face value as a signature of rotation for a highly inclined disk. Fig.~\ref{SiO-velocities}{\it b} nevertheless confirms the presence of non-circular motions in the CND: SiO emission is detected at regions of the p-v diagram which are {\it forbidden} by circular rotation: emission in quadrants I and II of Figure~\ref{SiO-velocities}{\it b}, can be interpreted by gas being in apparent counter-rotation and/or moving radially outward. In either case, this can be taken as evidence that gas orbits must be either elongated, if they lie in a common plane, or, alternatively, non-coplanar \footnote{in this context a combination of elongated and non-coplanar orbits cannot be excluded.}. Note that the detection of SiO gas in these regions cannot be explained by beam smearing effects: the deconvolved size of the emission for the velocity channels showing the largest excursions into I and II ($\sim$1155~km~s$^{-1}$ and 1125~km~s$^{-1}$, respectively)  is $\simeq$1.5$\times$ the beam size at $PA\sim$90$^{\circ}$. Similar features are detected in the  0.7$\arcsec$ resolution CO(2--1) map of S00 (Fig.4 of S00). As discussed in Sect.~\ref{Kinematics-CN}, CN emission is also associated with {\it forbidden} velocities. 

\begin{figure}[t!]
   \centering
   \includegraphics[width=7.5cm]{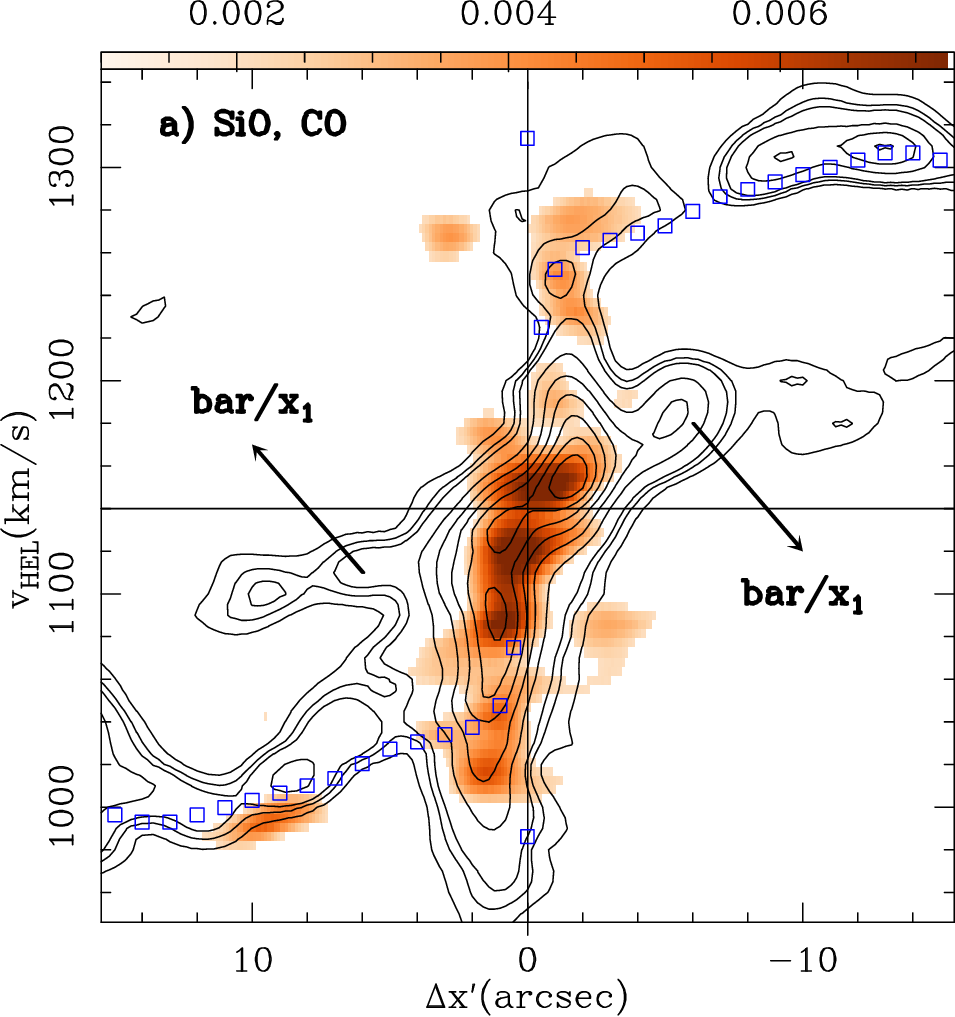}
   \includegraphics[width=7.5cm]{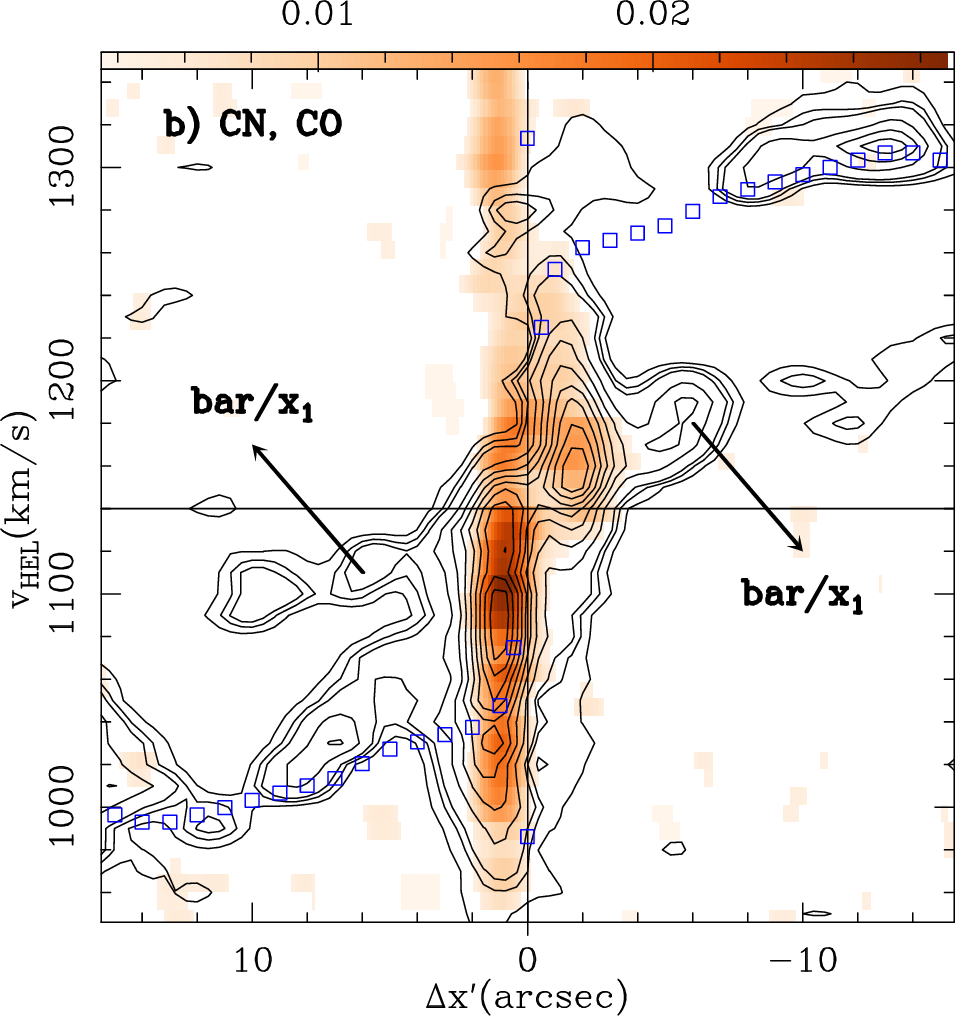}
   \caption{{\bf a)}~({\it Upper panel}) Position-velocity diagram of CO(1--0) (contours; data from S00 degraded to the resolution of SiO) and SiO(2--1) (color scale; this work) along $PA$=90$^{\circ}$. We highlight the kinematic signature of the bar leading edges, formed by the precession of x$_1$ orbits, seen in the {\it low} velocity CO component.  We plot the projected terminal velocities (empty squares) derived from the rotation curve fitted by S00. {\bf b)}~({\it Lower panel}) Same as {\bf a)} but showing the position-velocity diagram of CO(1--0) (contours; data from S00) and CN(2--1) (color scale; this work), degraded to the resolution of CO, along $PA$=90$^{\circ}$.}
              \label{SiO-CO-major}
\end{figure}

In the extended ILR region of the CND it is expected that the gas flow would trace the gradual transition from the $x_1$ orbits of the nuclear bar (between corotation and the outer ILR) to the $x_2$ orbits (between the ILRs) (e.g., Athanassoula~\cite{Ath92}; Buta \& Combes~\cite{But96}). Orbits of the  $x_2$  family would thus account for most of the CO and SiO emission detected at {\it high velocities} at $r\leq$2$\arcsec$ in the major axis p-v diagrams of Fig.~\ref{SiO-CO-major}{\it a}. Contrary to SiO emission, which is mainly restricted to the CND, CO emission is detected  over the SB ring ($\Delta$x'=$\pm$14$\arcsec$) but also in a region that connects the CND with the SB ring. A significant percentage of the CO emission on these intermediate scales ($\Delta$x'=$\pm$3--8$\arcsec$) is detected at {\it low velocities}, i.e., at velocities significantly below the terminal limit imposed by the rotation curve at this radii (Fig.~\ref{SiO-CO-major}{\it a}). This is the kinematic signature of the gas response to the stellar potential, characterized in this region by the crowding of $x_1$ orbits at the leading edges of the bar (S00; B00). As expected, the  $x_1$ + $x_2$  orbit combination produces a double-peaked line-of-sight velocity distribution in the CO p--v diagram that shows  the characteristic {\it figure-of-eight} trend as a function of radius where {\it high velocities} lie at small radii  (Kuijken \& Merrifield~\cite{Kui95}; Garc\'{\i}a-Burillo \& Gu\'elin ~\cite{Gar95}).


In the light of the published dynamical models of NGC\,1068 (S00; B00), we conclude that most of the SiO emission in the CND is associated with a region identified as the extended ILR region of the galaxy. Non-coplanar motions have alternatively been proposed to fit the CND kinematics.  Nonetheless in either case (nuclear bar or nuclear warp) the high velocity emission detected in SiO is physically associated with gas lying at small radii ($r\leq$2$\arcsec$). 


\begin{figure}[tb!]
   \centering
   \includegraphics[width=7cm]{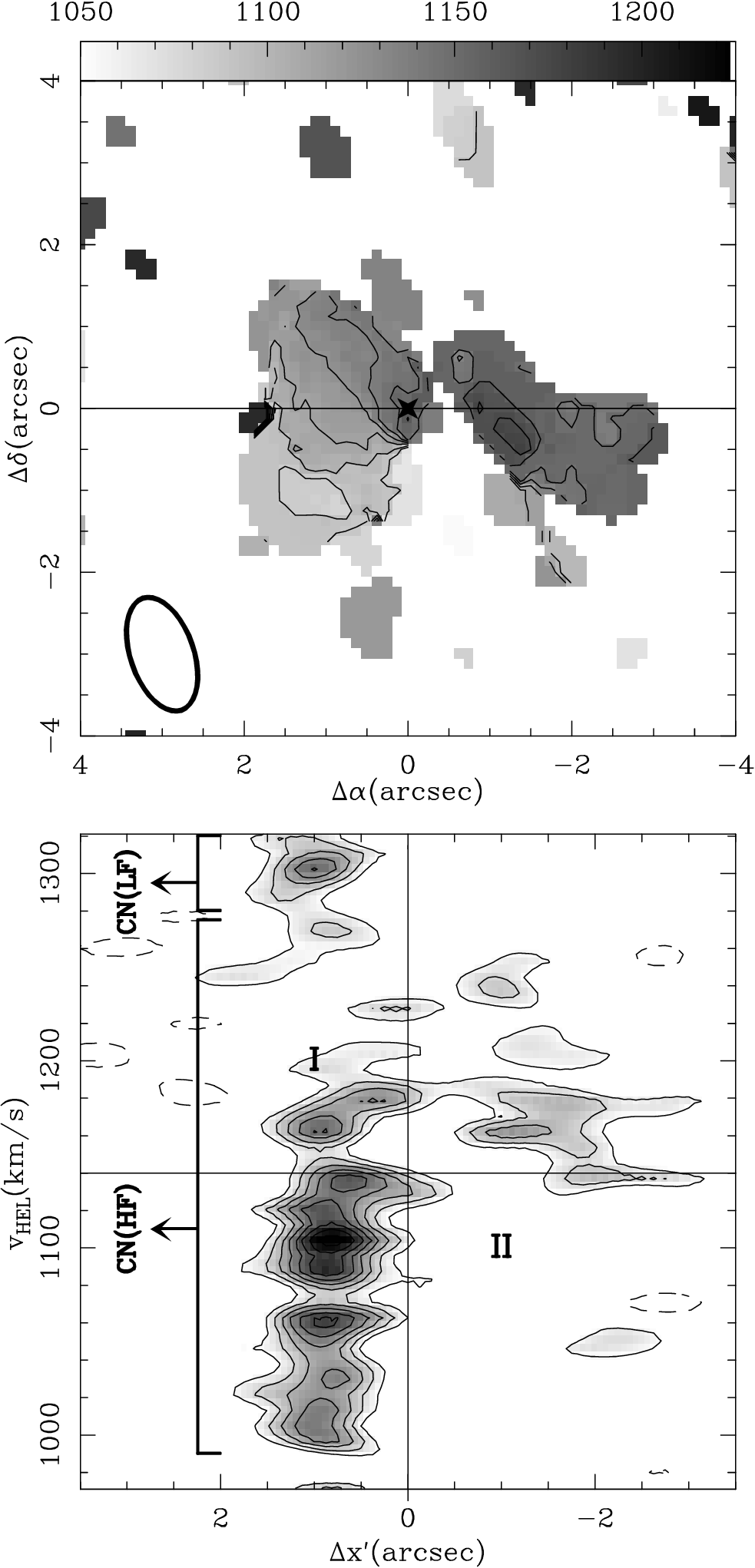}
   \caption{Same as Fig.~\ref{SiO-velocities} showing in {\bf a)}~({\it Upper panel}) CN isovelocities, derived from the high frequency (HF)  2--1 transition of CN, contoured over false-color velocity maps. Velocities span the range [1025~km~s$^{-1}$, 1200~km~s$^{-1}$] in steps of 25~km~s$^{-1}$. Similarly, we show in {\bf b)}~({\it Lower panel}) the  p--v diagram of CN along the major axis
   at $PA$=90$^{\circ}$. Contour levels are -15.2, 15.2, 22.8, to 68.4 in steps of 7.6~mJy~beam$^{-1}$. Velocities are referred to the HF 2--1 line. We highlight the range of the HF and LF lines of CN.}
              \label{CN-velocities}
\end{figure}

\subsection{CN kinematics}\label{Kinematics-CN}

Figure~\ref{CN-velocities}{\it a} shows the isovelocity contour map of the CND of NGC\,1068 derived from CN. Isovelocities have been obtained using a 3$\sigma$ clipping on the data. Compared to SiO, the higher spatial resolution of the CN maps provides a sharper but more complex picture of the gas dynamics in the CND. The rotating pattern of the CN emitting gas is highly perturbed. The overall kinematics suggest an east-west orientation of the major axis. However, at close sight the $PA$ of the major axis changes from 90$^{\circ}$ to 180$^{\circ}$ over the spatial extent of the E CN knot. In addition, isovelocities show an irregular pattern at the W CN knot. In spite of the more complex kinematics revealed by the higher resolution CN map on smaller scales, we will nevertheless assume in the following the same overall  orientation and kinematic parameters determined from SiO. As argued in Sect.~\ref{Kinematics-SiO}, this is a reasonable guess compatible with previous determinations based on the observed stellar and gas kinematics.     
 
Figure~\ref{CN-velocities}{\it b} shows the CN p-v diagram along $PA\sim$90$^{\circ}$. At this spatial resolution the p-v diagram shows a strong east-west asymmetry: most of the gas emission appears associated to the E knot. Similarly to SiO, CN emission is detected at regions of the p-v diagram which are {\it forbidden} by circular rotation. For CN, forbidden velocities appear mostly at quadrant I (see Fig.~\ref{CN-velocities}{\it b}).  
Due to the higher spatial resolution, the velocity gradient measured along the major axis is $\sim$a factor of two steeper in CN than in SiO. Otherwise,  the two molecular tracers of the CND share the same kinematic features, such as the existence of {\it high velocity} emission at small radii (for CN $r\leq$1$\arcsec$; see Fig.~\ref{CN-velocities}{\it b}), likely connected to  $x_2$  orbits. Like SiO, CN shows no detectable emission at {\it low velocities} on intermediate scales ($r\geq$3$\arcsec$; see Fig.~\ref{SiO-CO-major}~{\it b}).

\section{Molecular abundances}\label{$X$}

In Sect.~\ref{SiO-CO-global} we use the average SiO/CO and SiO/H$^{13}$CO$^{+}$ intensity ratios measured in the CND and the SB ring to derive the global abundance of SiO in the two regions, using a one-phase LVG radiative transfer model. We further analyze in Sect.~\ref{SiO-CN-CO-CND} the SiO/CO and CN/CO intensity ratio patterns inside the CND, as derived from the high-resolution PdBI maps. We discuss how these patterns can be interpreted in terms of different physical/chemical properties of molecular gas in the CND.


  \begin{figure}[tbp] 
     \centering
     \includegraphics[width=\hsize]{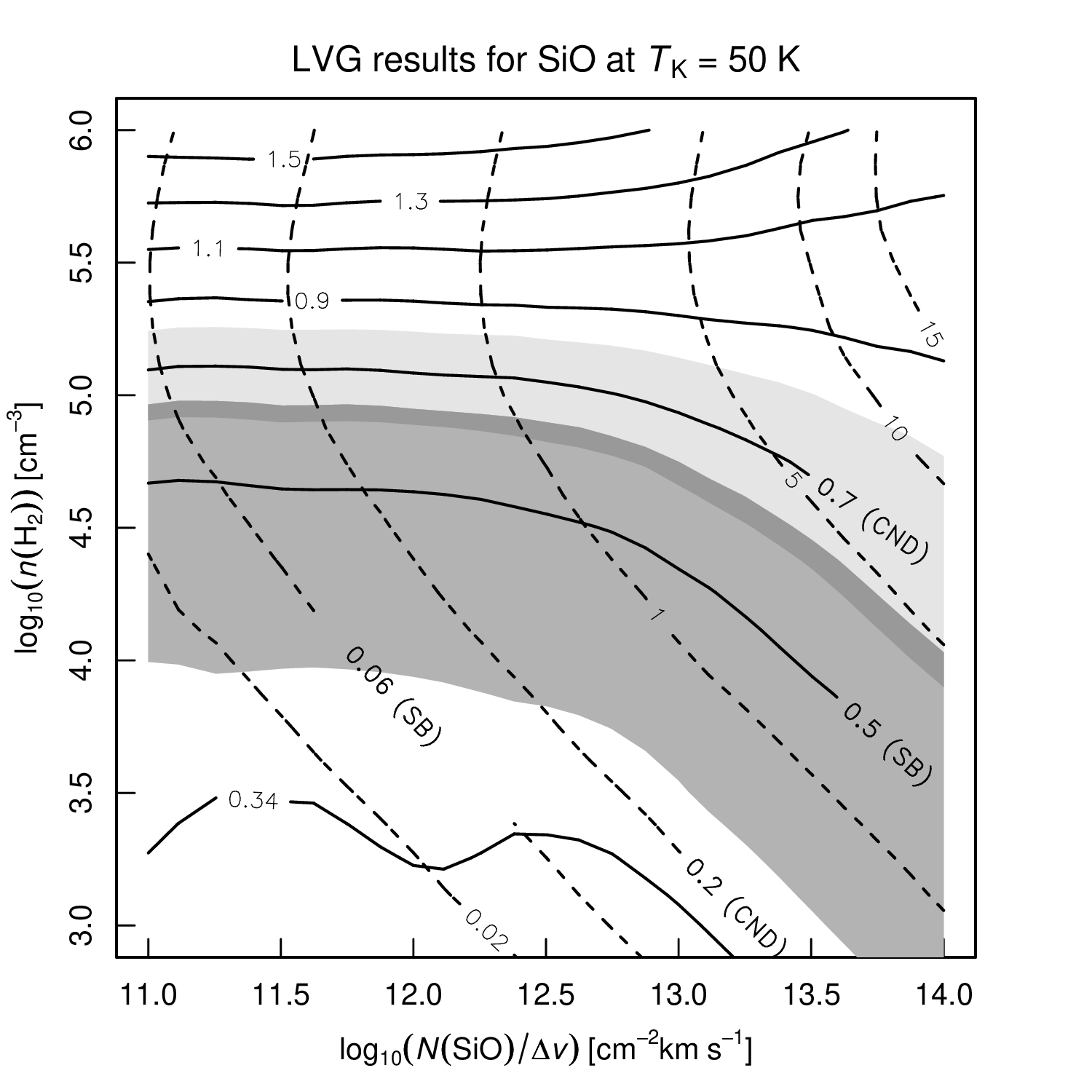} 
     \caption{LVG results for SiO, assuming a kinetic temperature of $T_\mathrm{K}=50$~K. 
     The $x$ and $y$ axes of the panel are, respectively, the SiO column density 
     per velocity width ($N_\mathrm{SiO}/\Delta\varv$) and the volume density of H$_2$
      molecules
     (n(H$_2$)) in a cloud. The solid and dashed lines indicate constant SiO(3--2)/SiO(2--1)$\equiv$$<R_{32/21}>$ line ratio and SiO(2--1) brightness temperature, respectively. The tags "SB" and 
     "CND" identify the mean
     values measured in the SB ring and the CND. The $<R_{32/21}>$ ratios within $1\sigma$ from the mean
      values in the CND and the SB correspond to the light- and medium-gray strips. The two strips
      overlap in a thinner band painted in darker gray. 
     }
     \label{fig-lvg1}
  \end{figure}



  \begin{figure}[tbp] 
     \centering
     \includegraphics[width=\hsize]{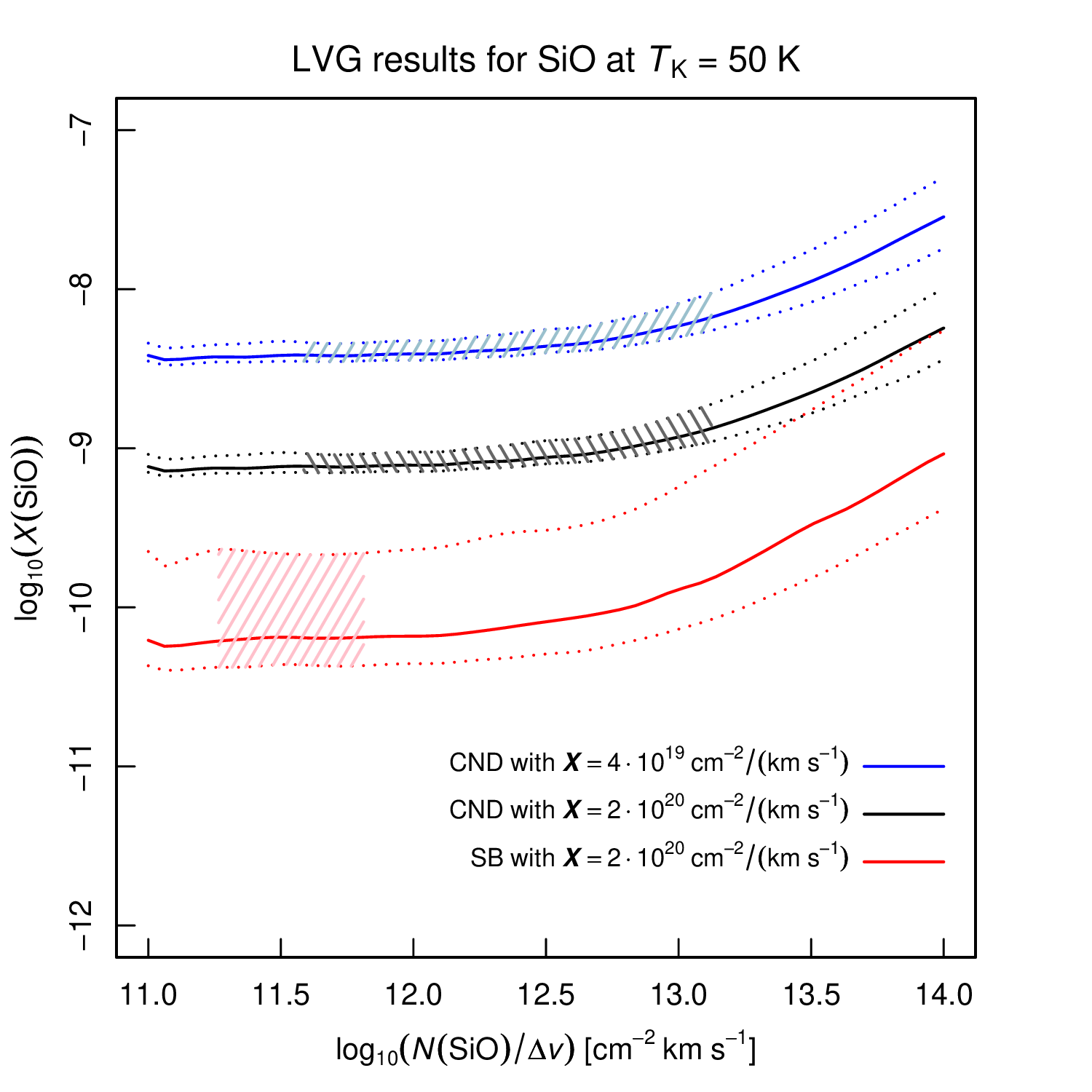} 
     \caption{SiO abundances estimated from the LVG model as a function of the 
     SiO column density per velocity width ($N_\mathrm{SiO}/\Delta\varv$) in a cloud 
     (see text for details). Each blue, black or red curve is defined by three parameters: 
     $<R_{32/21}>$, $<R_\mathrm{SiO/CO}>$ and ${\cal X}$. The conditions in the SB ring 
     are represented by the red curves and those in the CND by the others. The solid curves 
     correspond to the measured $<R_{32/21}>$ and the dotted ones to this value $\pm\sigma$  
     level. The SB solutions are only given for a Galactic ${\cal X}$, while in the case of the CND
      a five times lower ${\cal X}$ is also considered (blue). The areas where the most likely 
      solutions lie are hatched as explained in the text. They range in $N_\mathrm{SiO}/\Delta\varv$ they cover is so that  the CO(1--0) brightness temperature in a cloud is neither smaller than the measured brightness temperature, nor higher than the gas kinetic temperature (50~K).}
     \label{fig-lvg2}
  \end{figure}


\subsection{SiO abundances in the CND and the SB ring}\label{SiO-CO-global}

We have estimated the average SiO/CO velocity-integrated intensity ratios ($<R_\mathrm{SiO/CO} >$)  measured in the CND and the SB ring regions defined as follows. Based on the CO distribution shown in Fig.~\ref{SiO-maps}{\it a}, we delimit the SB ring as the region between $r\sim$10$\arcsec$ and $r\sim$20$\arcsec$ with SiO intensities $>$2.5$\sigma$ levels to derive integrated values. Similarly, the CND region is defined using a 2.5$\sigma$ clipping on the integrated intensities of the SiO map of Fig.~\ref{SiO-maps}{\it b}. Intensity units used to derive line ratios are K~km~s$^{-1}$ with K in T$_{mb}$ scale.  The $<R_\mathrm{SiO/CO} >$ ratios are thus equivalent to  brightness temperature ratios, as the line widths for SiO and CO are comparable at this spatial resolution. 

The maximum ratio in the SB ring, corrected by the different primary beam attenuation factors, is $\sim$0.004$\pm$0.001, a ratio estimated from the average spectrum of the clumps detected in SiO. This is a factor of $\sim$20 lower than the corresponding ratio averaged over the CND, estimated as $\sim$0.08$\pm$0.01.  The dichotomy between the CND and the SB ring is apparent from the remarkably different $<R_\mathrm{SiO/CO} >$ ratios measured in the two regions. The origin of such a dichotomy could be found in the different physical conditions or, alternatively, in the different chemical properties of molecular gas in these two regions. In the following, we use a radiative transfer code to fit the set of available line ratios in order to disentangle the contribution of different factors.   
 
 \subsubsection{LVG model results}\label{SiO-LVG}

To estimate the SiO abundances both in the CND and in the SB ring, we adopt a one-phase LVG approach to model the radiative transfer of the SiO emission.  We use an LVG code to translate the SiO brightness temperatures into the SiO column density per velocity width ($N_\mathrm{SiO}/\Delta\varv$) and the volume density of its collisional partner, H$_2$ (n(H$_2$)), in each cloud. 
We feed into the LVG code the average SiO(2--1) brightness temperature (from our PdBI observations) and the average SiO(3--2)/SiO(2--1)  velocity-integrated line ratio (from U04), hereafter $<R_{32/21}>$, measured in each target region. Provided that the 3--2 and 2--1 SiO emissions arise from the same clouds in each region, $<R_{32/21}>$  remains independent of the unknown filling factor $\eta_\mathrm{f}$.
To further constrain the model, the gas kinetic temperature is fixed at 50~K, as obtained by Tacconi et al.~(\cite{Tac94}) from multi-transition CO observations of the CND.     
This assumption is not critical, since the results barely depend on the gas temperature below, e.g., $100$~K. Quite likely, 100~K is a comfortable upper limit to the average temperature of the SiO emitting gas in a PdBI beam. 

The output of the LVG code is summarized in Fig.~\ref{fig-lvg1}. Since the filling factors are unknown the possible solutions for the CND (SB ring) clouds would lie over the corresponding $<R_{32/21}>$ isocontour, anywhere to the right of the intersect with the CND (SB ring) SiO(2--1) isocontour (i.e., the lower $\eta_\mathrm{f}$, the higher $N_\mathrm{SiO}/\Delta\varv$ in a cloud). The loci of possible solutions become two-dimensional bands, once we also consider the uncertainties in $<R_{32/21}>$. The measured $<R_{32/21}>$ ratios correspond to typical gas densities of $\sim10^5$~cm$^{-3}$ in the CND, while they are $\gtrsim5$ times lower in the SB ring.  

We search for a reasonable estimate of the SiO abundance with respect to H$_2$ ($X$(SiO)$\equiv N_\mathrm{SiO}/N_\mathrm{H_2}$) in the SB ring and the CND. For this we use ancillary CO observations of NGC~1068 to infer the H$_2$ column density. A full radiative transfer approach (e.g., LVG) is not possible, however, in the SB ring, as multi-transition CO observations are not available in this region. Alternatively, we use the standard conversion between the CO(1--0) integrated intensities and the $H_2$ column densities ($N_\mathrm{H_2}$) given by the so called  ${\cal X}$ factor in the two target regions.  The ${\cal X}$ conversion factor is of order 
$\sim 2\times10^{20}$~cm$^{-2}$/(km~s$^{-1}$) in the Milky Way (Strong et al.~\cite{Str88}), a value which roughly holds true for most nearby galaxies. For the SB ring, we adopt the Galactic ${\cal X}$ value. It is known, however, that a lower  ${\cal X}$ has been found in galaxy centers, where molecular clouds are subject to abnormal conditions (e.g., they are unlikely to be virialized). In particular, ${\cal X}$ in the CND of NGC~1068, could be $\sim5$ times lower than the standard value, according to the estimates of U04, a result which is in agreement with the findings obtained in the nuclear disks of other spiral galaxies (e.g., Garc\'{\i}a-Burillo et al.~\cite{Gar93}).

Under these assumptions, we proceed as follows to estimate $X$(SiO) in the SB ring and the CND:
\begin{itemize}
\item Each target region is characterized by fixed $<R_{32/21}>$  and $<R_\mathrm{SiO/CO} >$ values \footnote{as for $<R_{32/21}>$, we adopt for $<R_\mathrm{SiO/CO}>$ the average values measured in the SB and the CND.}, assumed to be independent of the filling factor, and by a certain ${\cal X}$. 
\item The measured $<R_{32/21}>$ allows us to calculate the SiO(2--1) brightness temperature as a function of $N_\mathrm{SiO}/\Delta\varv$ from the diagram in Fig.~\ref{fig-lvg1}.
\item $N_\mathrm{H_2}/\Delta\varv$ is proportional to the product of $<R_{\rm SiO/CO}>$ and the SiO(2--1) brightness temperature. Thus, ultimately, $X$(SiO) can be expressed in terms of a certain function of $N_\mathrm{SiO}/\Delta\varv$ parameterized by $<R_{32/21}>$, $<R_{\rm SiO/CO}>$ ratios and ${\cal X}$. 
\item  The range of values for $N_\mathrm{SiO}/\Delta\varv$ can be constrained. The lower limit to $N_\mathrm{SiO}/\Delta\varv$ corresponds to $\eta_\mathrm{f}=1$. On the other hand, for $\eta_\mathrm{f}$ arbitrarily low, the CO(1--0) brightness temperatures in a cloud would rise above the assumed kinetic temperature ($50$~K), which cannot occur if CO is collisionally excited. This sets an upper limit to $N_\mathrm{SiO}/\Delta\varv$.
\end{itemize}

The end product of this analysis is the plot in Fig.~\ref{fig-lvg2}. There we represent  $X$(SiO) as a function of $N_\mathrm{SiO}/\Delta\varv$  for the clouds in the SB ring and the CND. In the CND we explore two solutions with ${\cal X}$ equal to the standard factor and to a five times lower value. 
In either case, we note that the values of $X$(SiO) in the CND, $\sim$(1--5)$\times$10$^{-9}$, are significantly boosted with respect to those typically found in Galactic dark clouds ($\lesssim 10^{-12}$; e.g., Ziurys et al.~\cite{Ziu89}); this result calls for a mechanism of efficient dust grain processing in the CND. It is worth noting that the true SiO abundances could be even higher than our estimates if only a fraction of the CO emitting clouds were simultaneously emitting in SiO (the curves in Fig.~\ref{fig-lvg2} would have to be shifted upwards).   
We also note that $X$(SiO) is at least $\gtrsim$one order of magnitude higher in the CND than in the SB ring. Actually, were ${\cal X}$ $\sim$5 times lower in the CND than in the SB ring, the SiO abundances in the two target regions would differ by up to $\sim$two orders of magnitude.  


\begin{figure*}
  \centering  
   \includegraphics[width=8cm]{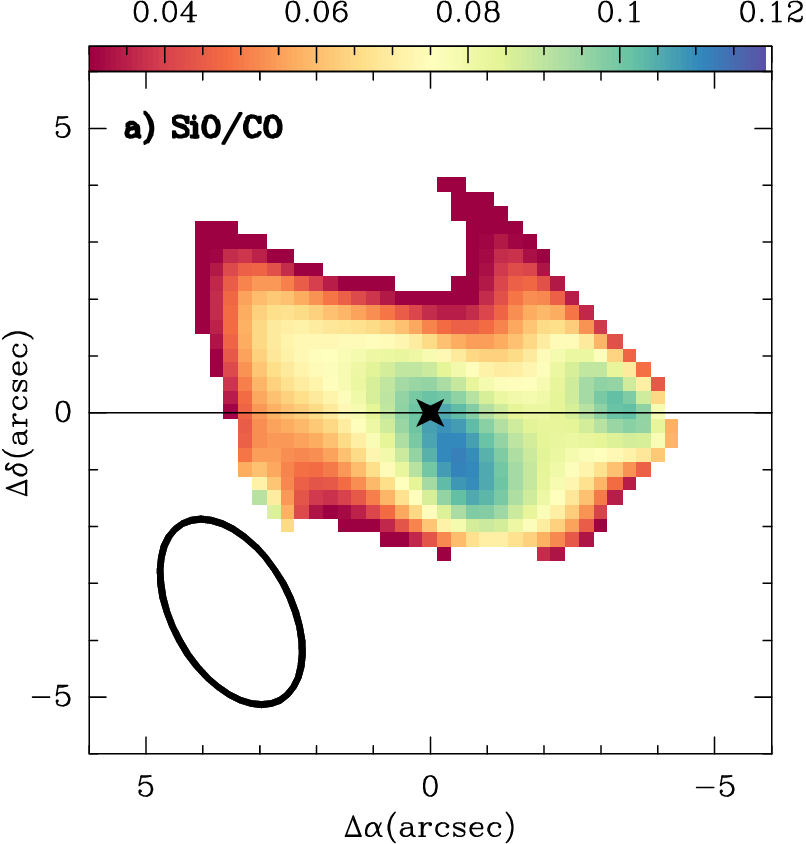}
   \includegraphics[width=8cm]{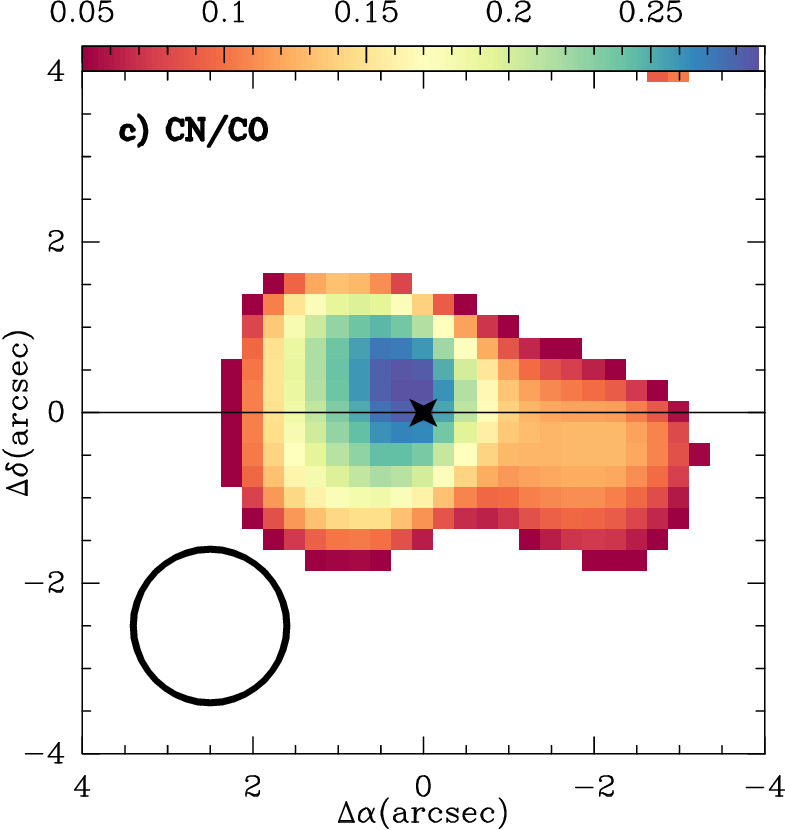}
   \includegraphics[width=8cm]{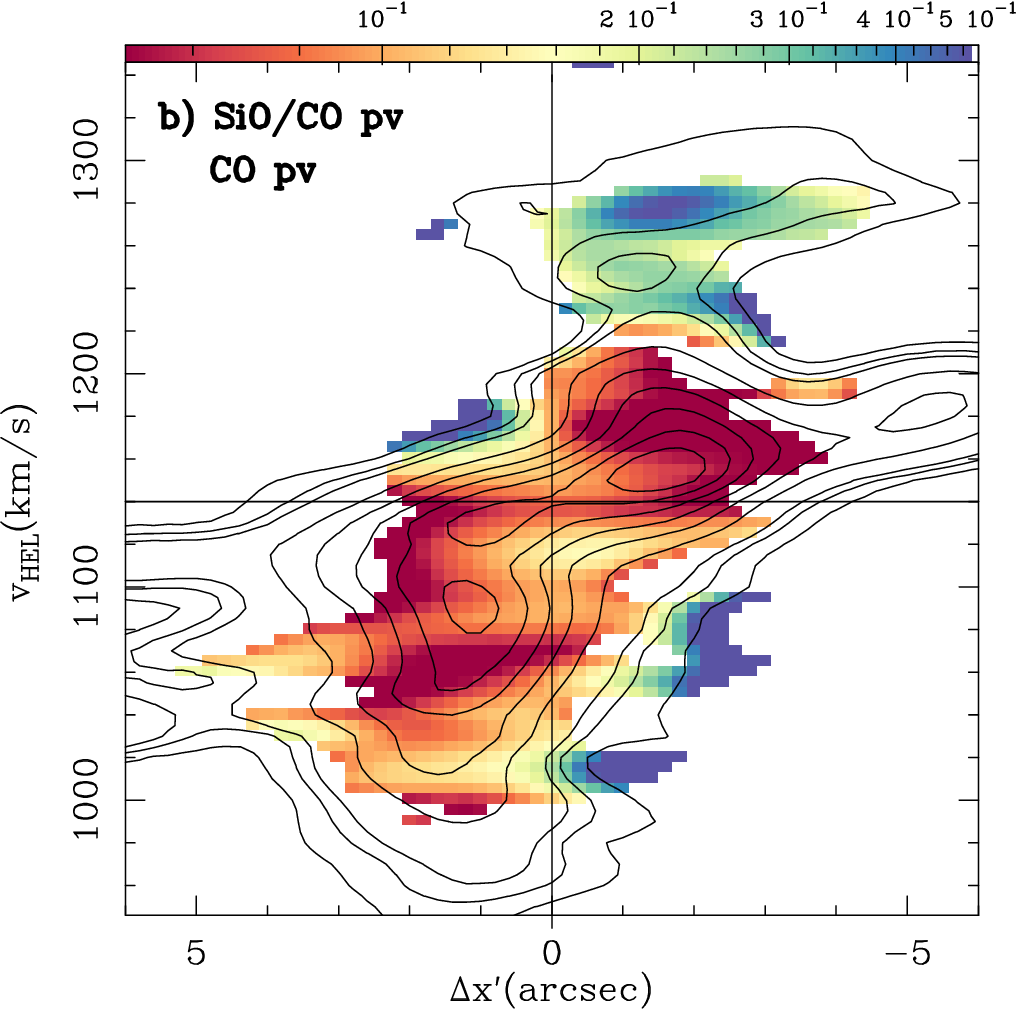}
   \includegraphics[width=8cm]{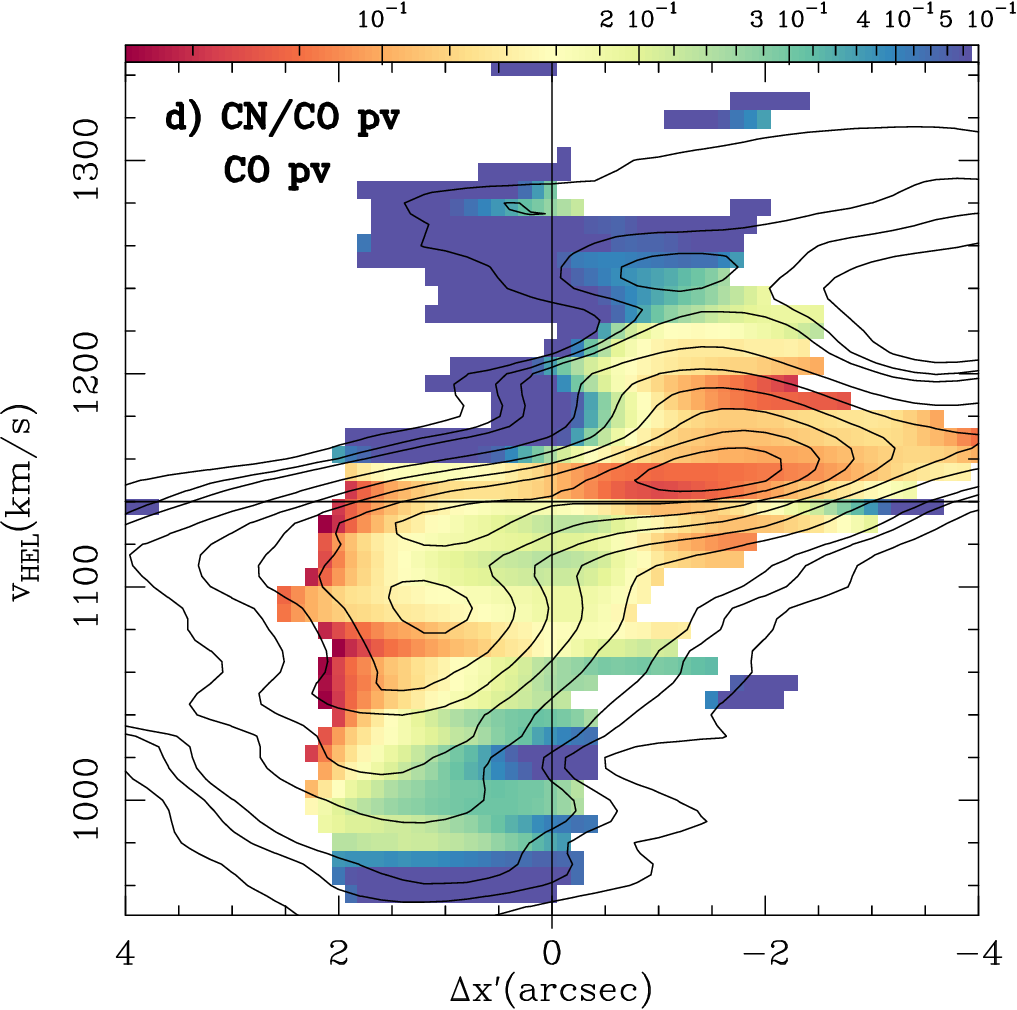}
        \caption{{\bf a)}~({\it Upper left panel})~The  SiO(2--1)/CO(1--0) velocity-integrated  intensity ratio ($R_\mathrm{SiO/CO}$) (color scale) derived at the spatial resolution of SiO inside the CND. This ratio is equivalent to a brightness temperature ratio for the observed common line widths of SiO and CO. An ellipse shows the SiO beam. {\bf b)}~({\it Lower left panel}) The SiO/CO ratio is here displayed as function of velocity  and spatial offset across the major axis at $PA$=90$^{\circ}$ (SiO/CO pv). The CO pv plot is displayed in contours (CO pv). {\bf c)} ({\it Upper right panel}) Same as {\bf a)} but showing here the CN(2--1)/CO(1--0) velocity-integrated intensity ratio ($R_\mathrm{CN/CO}$)(color scale) derived at the spatial resolution of CO. An ellipse shows the CO beam. {\bf d)}~({\it Lower right panel}) Same as {\bf b)} but adapted to compare here the CN/CO ratio along the major axis (CN/CO pv) with the corresponding CO pv plot.}
              \label{SiO-CN-CO-ratio}
\end{figure*}

\subsubsection{SiO/H$^{13}$CO$^+$ ratios}\label{SiOH13COp}

The results of the radiative transfer model described above can be compared to an alternative estimate of the abundance of SiO in the CND, 
based on the observed SiO/H$^{13}$CO$^+$ line ratio.  The H$^{13}$CO$^+$(1--0) line, observed simultaneously with the SiO(2--1) line, has also been detected in the CND of NGC~1068. Compared to SiO, the H$^{13}$CO$^+$ line is $\sim$a factor of three weaker. This ratio is significantly larger than that derived by U04 who reported a value $\sim$1 from a 30m spectrum of the CND. The disagreement between the PdBI and the 30m results is an indication that, compared to SiO, the contamination from the SB ring in H$^{13}$CO$^+$ is more severe, and cannot be accurately estimated using the lower resolution 30m map. 

Given the weakness of H$^{13}$CO$^+$ in the CND, we omit any discussion on the 2D morphology of the emission of this line.
We have nevertheless used H$^{13}$CO$^+$ to independently estimate the abundance of SiO in the CND using a LVG code to fit the global SiO(2--1)/H$^{13}$CO$^+$(1--0) intensity ratio measured in the nucleus of NGC~1068. The value of $X$(SiO) can be derived assuming a plausible range of physical conditions for the gas. This relies on two basic assumptions (see also discussion in Garcia-Burillo et al.~\cite{Gar00} and Usero et al.~\cite{Use06}). First, since SiO and H$^{13}$CO$^+$ have similar dipole moments and the observed transitions have comparable upper state energies, it is plausible to assume similar physical conditions for the two species. In our case we take the CND values estimated by U04 from the multi-line 30m survey of NGC~1068: n(H$_2$)$\sim$10$^{5}$cm$^{-3}$ and $T_{\rm K}=50$~K. As a second assumption, we take a standard abundance for H$^{13}$CO$^+$=2.5$\times10^{-10}$ and derive $X$(SiO) from the  SiO/H$^{13}$CO$^+$ column density ratio fitted by LVG. In our case this corresponds to a typical abundance of the main isotope X(H$^{12}$CO$^+$) =10$^{-8}$, and to an isotopic ratio [$^{12}$C]/[$^{13}$C]$\simeq40$, similar to the values found in the CND of our Galaxy (Wannier~\cite{Wan80}).
The derived abundance of SiO in the CND is $\sim$2.2$\times10^{-9}$. This estimate is comfortingly close to the lower limit of the $X$(SiO) value derived from the LVG modeling of the SiO/CO ratios described above. This agreement confirms that the one-phase LVG approach adopted in Sect.~\ref{SiO-LVG} provides a reasonable  estimate of the SiO molecular abundances in NGC~1068, in spite of the inherent limitations of such a simplified description of the molecular ISM in this model.    

 We thus conclude that the dichotomy between the CND and the SB ring, regarding the measured $<R_\mathrm{SiO/CO} >$ and $<R_{32/21}>$ ratios, cannot be attributed to a comparatively higher volume density of SiO gas in the CND, but mainly to a significant enhancement of the SiO abundance in this region.

\begin{figure}[tb!]
   \centering
   \includegraphics[width=7.5cm]{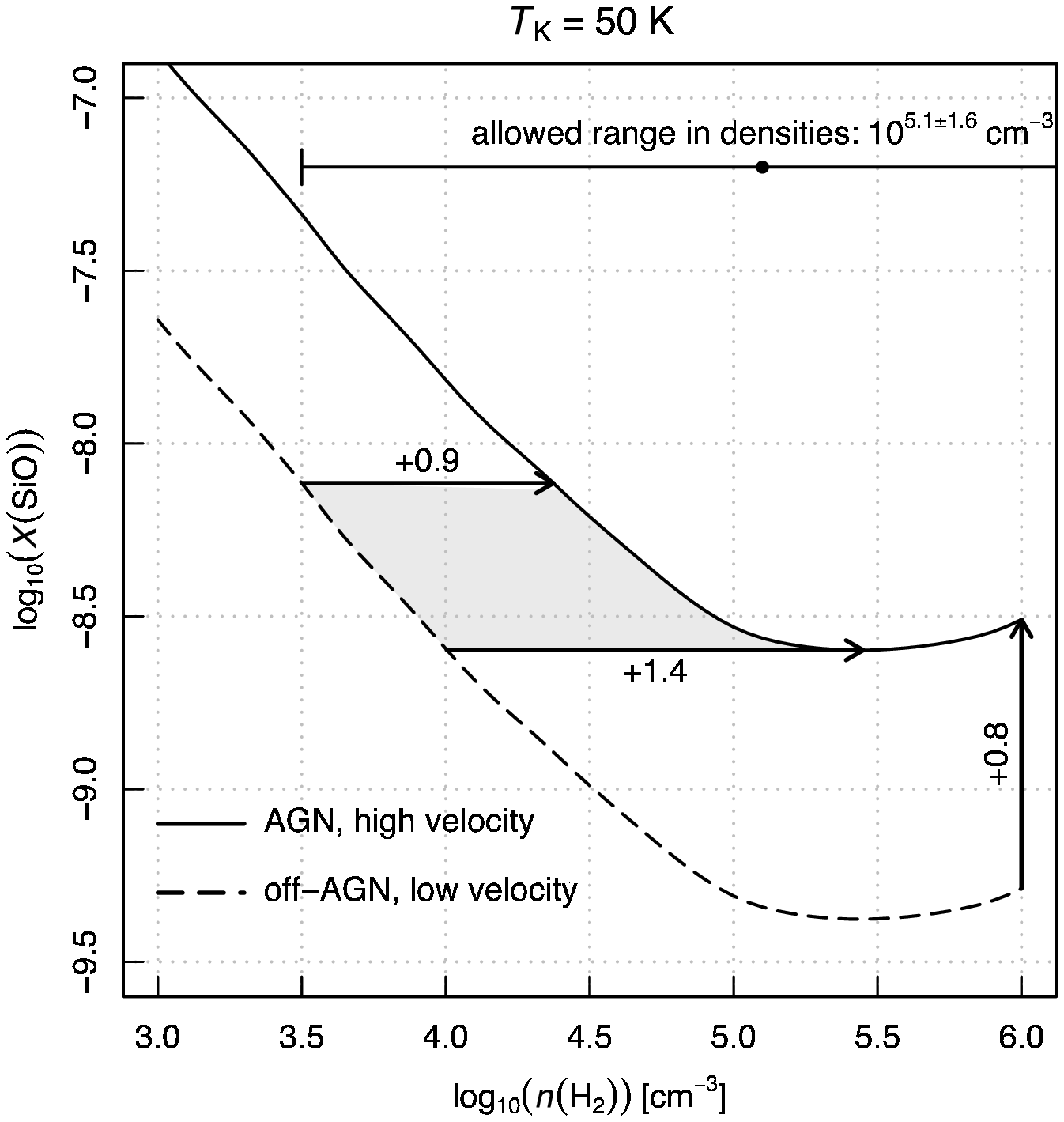}
   \includegraphics[width=7.5cm]{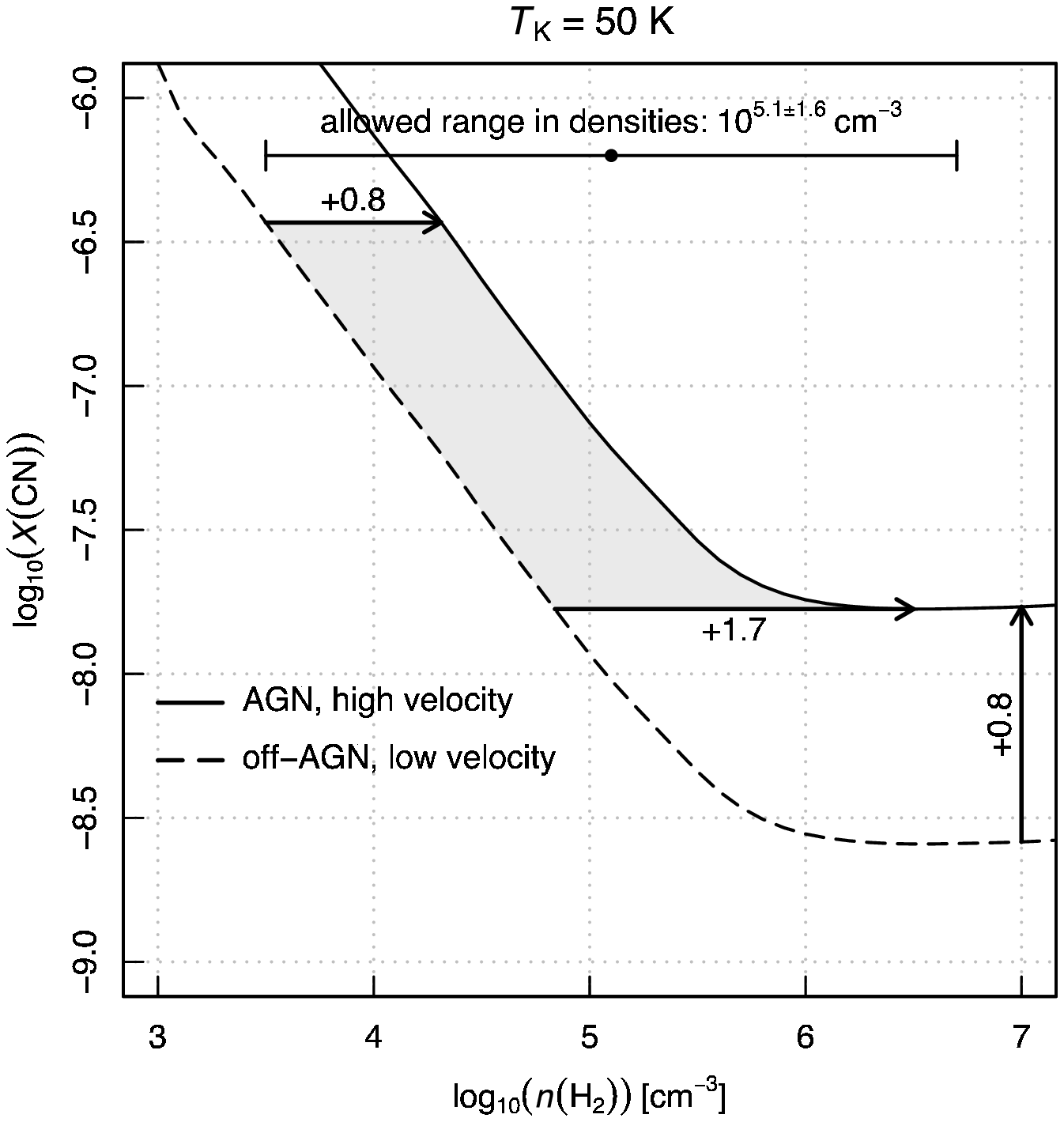}
   \caption{{\bf a)}~({\it Upper panel})~SiO abundance as a function of gas density in the {\em high velocity} gas at 
the position of the AGN (solid line) and in the {\em low velocity} at the edge of the CND 
(dashed line). The abundances are derived from the mean 
SiO and CO brightness temperatures within the respective velocity windows as defined in text. 
The SiO abundance in the two regions can be the same only 
within the grey area. {\bf b)}~({\it Lower panel}) Same as {\bf a)} but particularized for the density and abundance ranges for CN.}
              \label{SiO-CND-LVG}
\end{figure}
 
\begin{figure}[t!]
   \centering
   \includegraphics[width=6.5cm]{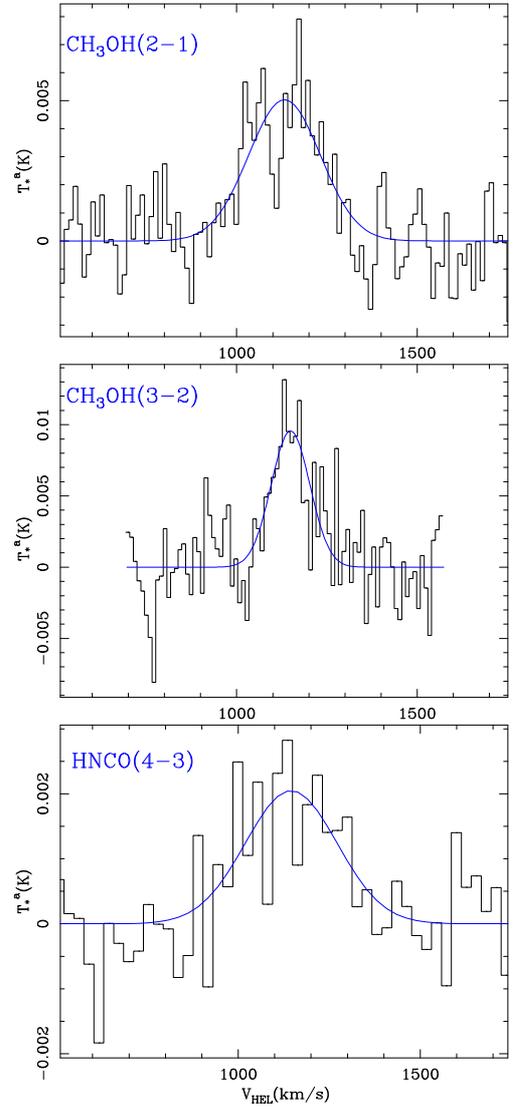}
   \caption{Spectra obtained with the IRAM 30m telescope in selected molecular line tracers of shock chemistry. We show in {\bf a)}~({\it Upper panel}) the spectra of the 2--1 line of CH$_3$OH. Intensity scale is in T$_a^*$ (K) units; velocities are in HEL scale (km~s$^{-1}$). The Gaussian fit to the spectrum is displayed in blue. {\bf b)}~({\it Middle panel})
Same as {\bf a)} but for the 3--2 line of CH$_3$OH.  {\bf c)}~({\it Lower panel})
Same as {\bf a)} but for the 4--3 line of HNCO.}
              \label{SiO-shocks}
\end{figure}
\subsection{Chemical differentiation in the CND}\label{SiO-CN-CO-CND}

\subsubsection{SiO abundances}\label{SiO-CND}

Figure~\ref{SiO-CN-CO-ratio}{\it a} shows the  SiO(2--1)/CO(1--0) velocity-integrated intensity ratio ($R_\mathrm{SiO/CO}$) in the CND, derived using a spatial resolution corresponding to that of SiO. The changes in $R_\mathrm{SiO/CO}$ are more significant along the major axis, where the spatial resolution is the highest. $R_\mathrm{SiO/CO}$ changes by $\sim$a factor of three inside the CND: it goes from a peak value of $\sim$0.11$\pm$0.01 in a region close to the AGN  to a minimum of $\sim$0.04$\pm$0.02 at the eastern boundary of the CND ($\Delta\alpha \sim$+3$\arcsec$); $R_\mathrm{SiO/CO}$ reaches a secondary maximum ($\sim$0.10$\pm$0.01) at the western edge of the CND ($\Delta\alpha \sim$--3$\arcsec$).  This range of values confirms and significantly expands the differences in the SiO/CO ratios between the east ($\sim$0.07) and the west ($\sim$0.10) lobes of the CND, identified by U04 as coming, respectively, from the blue and the red velocity components of the line profiles.  Figure.~\ref{SiO-CN-CO-ratio}{\it b} shows the SiO/CO ratio displayed here as a function of velocity and position across the major axis of the galaxy. The differences in the SiO/CO ratios, which go from $\sim$0.05 to $\sim$0.5, expand out to $\sim$one order of magnitude depending on the velocity channel and position along the major axis. Line ratios follow a regular pattern, where the highest ratios are associated with {\it high velocities}. In particular, the SiO/CO ratio is $\sim$0.35$\pm$0.05 at  $\mid v-v_{sys} \mid$=80--160km~s$^{-1}$ towards the AGN and the western secondary maximum. By contrast, the corresponding average ratios at {\it low velocities} ($\mid v-v_{sys} \mid <$+80~km~s$^{-1}$)  are $\sim$a factor of 6--7 lower ($\sim$0.05$\pm$0.01) at the edges of the CND ($\Delta x'$=$\pm$3$\arcsec$). 
 
The dichotomy between the SiO/CO ratios measured at {\it high} and {\it low velocities} can be interpreted in terms of different physical conditions (i.e., different n(H$_2$)), or, alternatively, as a signature of chemical differentiation (i.e., different $X$(SiO)) in the CND. Figure.~\ref{SiO-CND-LVG}{\it a} shows the range of possible LVG solutions that fit the 
CND ratios. We explore two {\it extreme} solutions: in the first case, a common $X$(SiO) is assumed, whereas in the second case, we adopt a common n(H$_2$). 
 While physical conditions and chemical abundances in the gas are to some extent interconnected, by exploring the two {\it extreme} cases described above, 
we can investigate which is the main driving cause of the reported changes in the line ratios. We restrict H$_2$ densities to lie within a range of values around the average density derived from the fit of the $<R_{32/21}>$ ratio measured in the CND (n(H$_2$)=10$^{5.1\pm1.6}$cm$^{-3}$). With these restrictions, Fig.~\ref{SiO-CND-LVG}{\it a} shows that we need  $\geq$one order of magnitude increase in n(H$_2$) to fit the progression of $R_\mathrm{SiO/CO}$
values observed from {\it low velocities} (at larger radii) to {\it high velocities} (at smaller radii). 
Alternatively, the scenario of a common n(H$_2$) but different $X$(SiO), suggests a much wider range of solutions for the CND. In this case, $X$(SiO) would need to be $\sim$a factor of 5--6 higher at {\it high velocities}.  Although a tenfold increase of n(H$_2$) cannot be formally excluded based on these data\footnote{A high-resolution $<R_{32/21}>$ map would be required to quantitatively probe SiO density changes inside the CND.}, we note that this solution implies that the change should occur on scales of $\sim$200--300~pc ($\sim$the CND radius). This is at odds with the significantly smaller differences ($\sim$a factor of 2--5) that exist between the average densities of SiO gas detected in the SB ring and the CND, two regions more than 1~kpc apart. 
Therefore we can interpret the spatial variations of  $<R_\mathrm{SiO/CO} >$ in the CND as the likely signature of chemical differentiation. In this scenario, the abundance of SiO  is significantly enhanced at {\it high velocities} (i.e., at small radii $r\leq$2$\arcsec$; see Sect.~\ref{Kinematics-SiO}). This interpretation is supported by the correlation found between  $<R_\mathrm{SiO/CO} >$ and the X-ray irradiation of the CND discussed in Sect.~\ref{XDR}.

\subsubsection{CN abundances}\label{CN-CND}
  
Figure~\ref{SiO-CN-CO-ratio}{\it c} shows the CN(2--1)/CO(1--0) velocity-integrated intensity ratio ($R_\mathrm{CN/CO}$) derived in the CND, using a spatial resolution corresponding to that of the CO map.  The average $R_\mathrm{CN/CO}$ ratio in the CND is $\sim$0.15.
The changes in $R_\mathrm{CN/CO}$ shown in Fig.~\ref{SiO-CN-CO-ratio}{\it c}  are significant along the major axis. $R_\mathrm{CN/CO}$ changes by $\sim$a factor of three inside the CND. Similarly to the SiO/CO ratio, the maximum CN/CO value ($\sim$0.3$\pm$0.03) corresponds to a region very close to the AGN. $R_\mathrm{CN/CO}$ decreases to $\leq$0.1$\pm$0.04 at the edge of the CND.
Figure.~\ref{SiO-CN-CO-ratio}{\it d} shows the CN/CO ratio here displayed as a function of velocity and position across the major axis of the galaxy. Similarly to the case of SiO discussed above, an inspection of Fig.~\ref{SiO-CN-CO-ratio}{\it d} shows that the differences in the measured CN/CO ratios are boosted to reach $\sim$one order of magnitude depending on the velocity channel and position along the major axis. Line ratios follow a regular pattern, with the highest ratios being associated with {\it high velocities}. Towards the AGN, the highest CN/CO values ($\sim$0.7$\pm$0.09) correspond to the {\it high velocity} component of the emission. This is in clear contrast with the corresponding values measured at {\it low velocities} ($\sim$0.1$\pm$0.02).
Following the scheme of Sect.~\ref{SiO-CND}, we have explored two families of LVG solutions that fit the range of CN/CO ratios of the CND assuming a common  $X$(CN) or a common n(H$_2$).  For simplicity we restrict H$_2$ densities to lie within the same range values explored for SiO. P\'erez-Beaupuits et al.~(\cite{Per09}), based on a multi-transition CN analysis, have derived a more restricted density range (n(H$_2$)=10$^{6\pm1}$cm$^{-3}$), which nevertheless lies within the range explored here.     
In the first scenario we need about an order of magnitude increase in n(H$_2$) to fit the progression of ratios from {\it low-velocities} to {\it high-velocities} (see Fig.~\ref{SiO-CND-LVG}{\it b}). The corresponding range of solutions in the second scenario requires  $\sim$a factor of 5--6 increase in $X$(CN).  For reasons similar to those described in Sect.~\ref{SiO-CND}, we favor an interpretation of the spatial variations of  $<R_\mathrm{CN/CO} >$ in the CND in terms of chemical differentiation. The abundance of CN  would be significantly enhanced at {\it high velocities} (i.e., at small radii $r\leq$1$\arcsec$; see Sect.~\ref{Kinematics-CN}). This interpretation is also supported by the correlation found between  $<R_\mathrm{CN/CO} >$ and the X-ray irradiation of the CND discussed in Sect.~\ref{XDR}.

\begin{table*}[!htp]
\begin{center}
\caption{Line ratios measured in prototypical regions.}    
{
{
\begin{tabular}{lcccccc}
\hline
\hline
\noalign{\smallskip}
Source	& ${CH_3OH(2-1)\over SiO(2-1)}$ & ${CH_3OH(3-2)\over SiO(3-2)}$& ${HNCO(4-3)\over SiO(2-1)}$& ${SiO(2-1)\over CN(2-1)}$& ${CN(2-1)\over HCN(1-0)}$ & References\\ 
\noalign{\smallskip}
\hline
\noalign{\smallskip}
NGC1068-CND		&2.3$^*$~(1.8)  & 2.1$^*$~(1.7) &  1.2$^*$~(1.2) & 0.9 & 0.25 & 1, 2 \\
\noalign{\smallskip}
\hline
\noalign{\smallskip}
L1157-B1		& 2$^*$~(1.7) & 1.9$^*$~(1.7) & -- & 3$^*$~(7.5) & 0.11$^*$~(0.04) & 3, 4  \\
L1157-B2		& 2.3$^*$~(2.1) & 2.9$^*$~(2.7) & -- & $>$7$^*$~($>$12) & $<$0.04$^*$~($<$0.03) & 3, 4  \\
\noalign{\smallskip}
\hline
\noalign{\smallskip}
IC342-Ndisk 		&-- & 4.5$^*$~(3.6) & -- & -- & --& 5, 6   \\
IC342-N arm 	& 5 & -- & 3.5 & -- & -- & 7, 8  \\
\noalign{\smallskip}
\hline
\noalign{\smallskip}

\end{tabular}

\tablefoot{We list line ratios derived in the CND of NGC~1068 (NGC1068-CND), the B1 and B1 bullets of the young stellar object L1157, as well as the nuclear disk 
and Northern spiral arm of the nearby galaxy IC342 (IC342-Ndisk and IC342-N arm, respectively). Line ratios are given in $T_{mb}$ units; these are corrected by source coupling factors for single-dish observations. Uncorrected ratios are identified by an asterisk; corrected ratios (in $T_b$-{\it source} units) are given in parentheses when necessary.}

\tablebib{(1)~ U04; (2)~this work; (3)~Nisini et al.~(\cite{Nis07}); (4)~P\'erez-Guti\'errez~(\cite{Per99}); (5)~H\"uttemeister et al.~(\cite{Hue97}); (6)~Garc\'{\i}a-Burillo \& Mart\'{\i}n-Pintado~(\cite{GarMP01}); (7)~Meier \& Turner~(\cite{Mei05}); (8)~Usero et al.~(\cite{Use06}).}
}

}
\end{center}
\end{table*}
%

\section{Molecular gas chemistry in the CND of NGC~1068}\label{Models}

In the following we consider the pros and cons of two different scenarios,  shocks and XDR chemistry, regarding their ability to explain the molecular abundances measured in the CND of NGC~1068 for SiO and CN. Complementary information provided by other molecular tracers is a key for this discussion. Furthermore, we use the information extracted from the observed kinematics of the gas and the relation of the derived molecular abundances with the X ray irradiation in the CND. 


\begin{figure*}
   \centering
   \includegraphics[width=17cm]{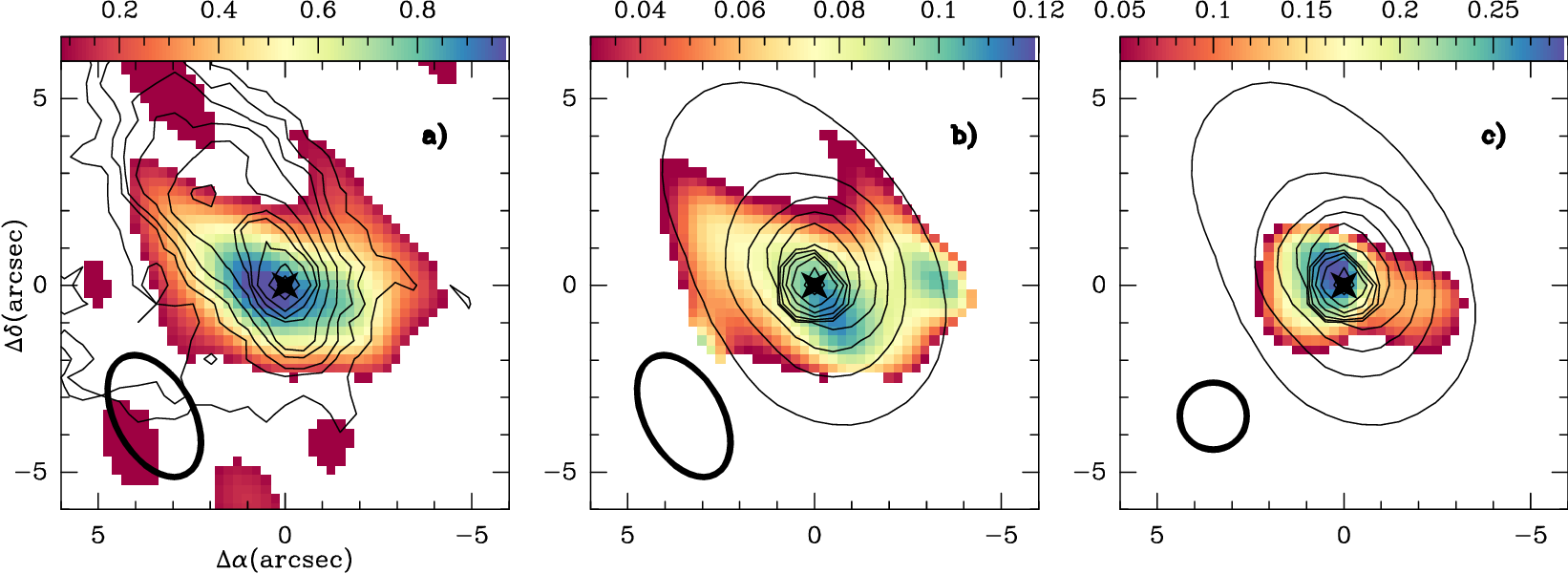}
    \caption{{\bf a)}~({\it Left panel}) The Chandra X-ray image of NGC~1068 (contours:15, 30, 50, 100, 200, 400, 600, 1000, 2000 and 3000 counts) obtained in the 0.25--7.5~keV band by Young et al.~(\cite{You01}) is overlaid  on the PdBI SiO map (color scale in units of Jy~km~s$^{-1}$~beam$^{-1}$). {\bf b)}~({\it Middle panel}) The X-ray image obtained in the 6--8~keV band by Ogle et al.~(\cite{Ogl03}) (contours: 0.2, 0.5, 1, 2, 4, 8, 15, 25, 40, 80 and 100 counts) is overlaid on the SiO(2--1)/CO(1--0) brightness temperature ratio (color scale) at the SiO spatial resolution. {\bf c)}~({\it Right panel}) Same as {\bf b)} but with the X-ray image obtained in the 6--8~keV band overlaid on the CN(2--1)/CO(1--0) ratio (color scale) at the CO spatial resolution. Ellipses show beams of SiO and CO as in Fig.~\ref{SiO-CN-CO-ratio}.
}
              \label{SiO-CN-Xraystot}
\end{figure*}

\subsection{Shock chemistry}\label{Shocks}

To explore the prevalence of shock chemistry in the CND of NGC~1068 we have used the IRAM 30m telescope to observe a set of lines of two molecular tracers of shocks:  CH$_3$OH and HNCO.  We discuss below the line ratios derived for CH$_3$OH, HNCO, SiO and CN in NGC~1068, and compare these with the ratios derived in galactic and extragalactic templates of shock chemistry. We also analyze the potential drivers of shocks in the CND. 

\subsubsection{Tracers of shocks: CH$_3$OH and HNCO}\label{tracers-shocks}

On theoretical grounds, large abundances of methanol can only be produced in gas-phase via evaporation and/or disruption of icy mantles. {\it  Fast} shocks ($v_{shock}>$15--20~km~s$^{-1}$) can destroy
the grain cores, liberating refractory elements, like Si, to the gas phase (Caselli et
al. \cite{Cas97}; Schilke et al. \cite{Sch97}). By contrast, {\it slow} shocks ($v_{shock}<$10--15~km~s$^{-1}$) are able to process the icy grain mantles, but not the grain cores (Millar et al.~\cite{Mil91}; Charnley et al.~\cite{Cha95}). The different location of
Si-bearing material (cores) and solid-phase CH$_3$OH (mantles) in dust grains implies that SiO and  CH$_3$OH are good
tracers of {\it fast} and {\it slow} shocks, respectively. It is expected that the disruption of dust grains by {\it slow} shocks can inject grain mantle material into the molecular ISM without destroying the molecules  (Bergin et al.~\cite{Ber98}, Mart\'{\i}n-Pintado et al~\cite{Mar01}). In particular, the abundance of methanol is seen to be enhanced in shocks associated with molecular outflows by more than two orders of magnitude over the values typically derived in cold molecular clouds (e.g., Bachiller \& P\'erez-Guti\'errez~\cite{Bac97}; P\'erez-Guti\'errez~\cite{Per99}). In external galaxies 
Meier \& Turner~(\cite{Mei05}) associate over-luminous methanol lines to shocks in IC~342. The close association between SiO and CH$_3$OH emission in IC~342, discussed by Usero et al.~(\cite{Use06}) corroborates this picture.


Although it is still debated what is the main production mechanism of HNCO, models involving dust grain chemistry seem to be the most successful at increasing the abundances of HNCO in gas phase.  There is also supporting observational evidence that HNCO is related to shocks. In particular, Zinchenko et al.~(\cite{Zin00}) finds a good correlation between HNCO and SiO lines in dense cores of our Galaxy. 
Mart\'{\i}n et al.~(\cite{Mar08}) have presented evidence of enhanced HNCO in molecular clouds suspected to suffer shocks in the Galactic Center. On larger scales, Meier \& Turner~(\cite{Mei05}) interpret the large abundance of HNCO measured in the nuclear spiral structure of IC~342 as due to molecular shocks. In a later study, Mart{\'{\i}}n et al.~(\cite{Mar09}) observed several lines of HNCO in a sample of nearby starbursts and AGNs, concluding that molecular shocks can enhance the abundance of HNCO in galaxies. More recently, Rodr\'{\i}guez-Fern\' andez et al.~(\cite{Rod10}) have reported a significant enhancement of HNCO in the bipolar outflow L1157, related to the shock chemistry at work in this prototypical young stellar object.  

\subsubsection{Tracers of shocks in NGC~1068}\label{30m-data}

Figure~\ref{SiO-shocks} shows the spectra obtained in the 3--2 and 2--1 line of CH$_3$OH and in the 4--3 line of HNCO towards the CND of NGC~1068 with the IRAM 30m telescope. Emission is detected in the three lines at significant levels (S/N ratio $\sim$7--12). Within the errors, the three line profiles appear centered around v$_{sys}$. However, the 3--2 line of methanol is a factor of 2 narrower than the other two lines. 
Such a difference cannot be attributed to the $\sim$1.5$\times$ smaller beam of the CH$_3$OH(3--2) line ($\sim$17$\arcsec$), as the maximum spread of velocities is already reached on the scales of the CND. By contrast this result reflects the lower excitation of methanol lines at {\it high-velocities} (likely arising in the inner CND, as shown by the SiO and CN interferometer maps). This supports the view that there is no significant increase of molecular densities at small radii on the scales of the CND (Sects.~\ref{SiO-CND} and ~\ref{CN-CND}).

Table~2 shows the average CH$_3$OH(2--1)/SiO(2--1), CH$_3$OH(3--2)/SiO(3--2),
HNCO(4--3)/SiO(2--1),  SiO(2--1)/CN(2--1)  and CN(2--1)/HCN(1--0)  line ratios in the CND of NGC~1068 in T$_{mb}$ scale.  We have corrected line ratios from beam dilution when necessary (single-dish observations) using the usual prescription for a point-like source, which is applicable in this context for the CND.  Line ratios obtained in NGC~1068 are compared with results obtained in the molecular lobes (known as B1 and B2) of the bipolar outflow L1157 (data taken from P\'erez-Guti\'errez~\cite{Per99}  and Nisini et al.~\cite{Nis07}). We also list, when available, the line ratios measured in the face-on barred galaxy  IC~342 (data taken from H\"uttemeister et al.~\cite{Hue97},  Meier \& Turner~\cite{Mei05} and  Usero et al.~\cite{Use06}).  In the case of L1157 we have applied a correction for beam dilution assuming that the sizes of molecular emission in B1 and B2  are 18$\arcsec$ and 30$\arcsec$, respectively.

As a result of this comparison we note the similarity between the ratios derived from CH$_3$OH, HNCO and SiO lines in L1157 and NGC~1068.
At face value, this indicates that most of the SiO emission in the CND of NGC~1068 could be explained by {\it fast} shocks similar to those identified in the young bipolar outflow L1157. A comparison between the SiO/CN and CN/HCN ratios measured in L1157 and  NGC~1068, indicates a remarkable excess of CN emission in NGC~1068, however. L1157 is the only outflow in which CN(2--1) emission has been detected in the high velocity gas associated with the bow shocks. But even in the case of this extremely molecular-rich young outflow, the SiO/CN intensity ratio is $\sim$a factor of 10 higher than in the CND of NGC~1068. This CN excess questions the suitability of the bipolar outflow template in NGC~1068.   The age dating of the most recent star formation episode of the CND ($>$2--3$\times$10$^8$~yrs old; Davies et al.~\cite{Dav09}) also questions the hypothesis that shocks in NGC~1068 can be interpreted as stemming from an embedded population of {\it young} stellar objects.

Alternatively, large-scale shocks produced by cloud-cloud collisions, which are enhanced due to the complex orbital dynamics in the ILR region of the CND, could be a viable mechanism to explain the emission of SiO, CH$_3$OH and HNCO in NGC~1068.  SiO maps show the existence of  perturbed kinematics of molecular gas in the CND (Sect.~\ref{Kinematics-SiO}). The prevalence of SiO emission at {\it high} and {\it forbidden velocities} implies enhanced X(SiO) in gas following non-circular orbits. Furthermore, the line ratios discussed in Sect.~\ref{SiO-CND} suggest an enhancement of SiO abundances at the extreme velocities identified in the CND.  A similar link between density waves and large-scale molecular shocks has been proposed to explain the enhancement of SiO in the circum-nuclear disks of our Galaxy (Mart{\'{\i}}n-Pintado et al.~\cite{Mar97}; H\"uttemeister et al.~\cite{Hue98}; Rodr{\'{\i}}guez-Fern{\'a}ndez et al.~\cite{Rod06}) and of the starburst galaxies IC~342 and NGC~253 (Usero et al.~\cite{Use06}; Garc{\'{\i}}a-Burillo et al.~\cite{Gar00}). Usero et al.~(\cite{Use06}) analyzed the large-scale molecular shocks produced in the ILR region of IC~342. Shocks in this galaxy seem to arise during cloud-cloud collisions at low velocities, only after the kinetic energy associated with the density wave driven streaming motions has partly dissipated into turbulence. The comparison between 
IC~342 and NGC~1068 indicates that CH$_3$OH/SiO and HNCO/SiO ratios are $\sim$a factor of two to three lower in NGC~1068. The detection of  over-luminous SiO lines in NGC~1068 could be the signature of comparatively higher velocity shocks in this galaxy, possibly produced by mechanisms unrelated to density waves. Other scenarios designed to fit the CND kinematics, in particular those invoking a jet-ISM interaction, which would give rise to a nuclear warp and, eventually, to an expanding ring, could produce {\it faster} molecular shocks in NGC~1068. 

Gallimore et al.~(\cite{Gal96}) analyzed the morphology and spectral index of the NGC~1068 jet and concluded that the radio plasma may have been diverted at a shock interface with the molecular ISM close to the AGN  (knot {\it C} at $r\sim$30-40pc). Molecular shocks in the CND could have propagated around the jet-ISM working surface. The H$_2$ line map published by Davies et al.~(\cite{Dav08}; see also M\"uller-S{\'a}nchez et al.~\cite{Mue09}) reveals a strongly perturbed velocity field at a region located just 0.5--1$\arcsec$ (35--70~pc) north of the AGN.  The enhanced SiO abundances derived at high velocities 'close' to the AGN may well be the signature of this jet-ISM interaction.  Nevertheless, the fact that SiO gas is detected elsewhere in the CND suggests that shocks would be at work on larger scales, either driven by an expanding ring or by the perturbed kinematics at the ILR region.   However, whereas shock chemistry can explain the strength and the line ratios of the different molecular shock tracers detected in NGC~1068, it is not a valid explanation for the measured CN abundances.

The discrepancies mentioned above motivate the search of a framework different to shock chemistry that would be able to simultaneously account  for the high molecular abundances of SiO and CN measured in NGC~1068 (Sect.~\ref{XDR}).


\begin{figure*}
   \centering
   \includegraphics[width=12cm]{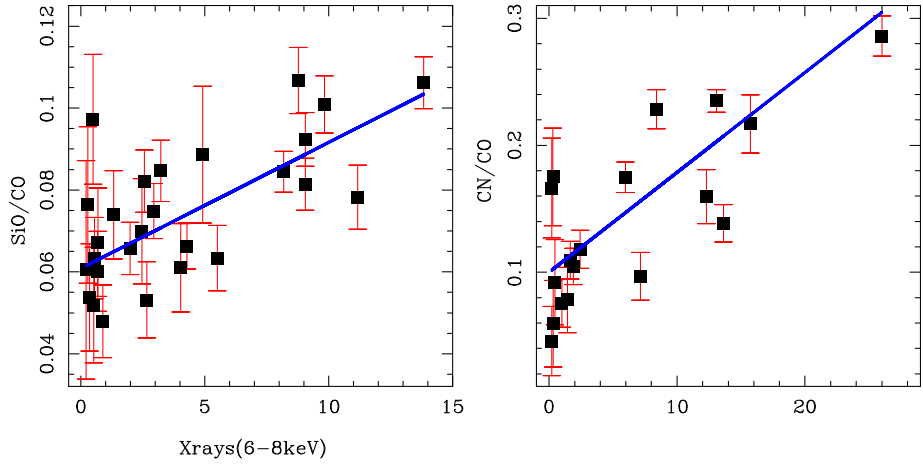}
    \caption{{\bf a)}~({\it Left panel}) The SiO(2--1)/CO(1--0) brightness temperature ratio versus the Xray flux in the 6--8~keV band. The straight line represents  the best fit to the points taking into account uncertainties visualized by the errorbars. The coefficient of determination of the regression is $r^2=0.6$.  {\bf b)}~({\it Right panel}) Same as {\bf a)}  but showing the CN(2--1)/CO(1--0) ratio. The coefficient of determination of the regression is $r^2=0.7$. 
 }
              \label{Xrays-correlations}
\end{figure*}

\subsection{X-ray chemistry}\label{XDR}

We describe below the main properties of X-ray emission in NGC~1068, with a particular emphasis on the morphology, energetics and origin of hard X-ray emission in the CND (Sect.~\ref{Xrays}). We further discuss the observational evidence for X-ray chemistry in the nucleus of  NGC~1068, and how the SiO and CN data 
specifically fit into this picture (Sect.~\ref{XDRchemistry}).   

\subsubsection{X-ray emission in NGC~1068}\label{Xrays}

We superpose in Fig.~\ref{SiO-CN-Xraystot}{\it a} the subarcsec resolution X-ray image of NGC~1068 obtained in the 0.25-7.50~keV band with Chandra (contours) on the SiO map (color scale). The X-ray image has been taken from the 3.2s frame time data published by Young et al.~(\cite{You01}). The 0.25-7.50~keV band image incorporates the contribution from soft to hard X-rays and includes both continuum and line emission. The morphology of the X-ray image reveals spatially resolved emission distributed in a compact yet spatially resolved component located on the AGN, and an elongated source that extends to the northeast, similarly to the inner radio jet of the galaxy.  The northeast elongation of X-ray emission contrasts with the overall east-west elongation seen in the CND molecular material detected in SiO and CN.  Lower level X-ray emission is detected outside the  inner 12$\arcsec \times$12$\arcsec$ field of view (FOV) displayed in  Fig.~\ref{SiO-CN-Xraystot}{\it a}. This outer emission takes the shape of a complex spiral-like structure that  extends out to $r\sim$30-40$\arcsec$ (see Young et al.~\cite{You01} for a detailed description).
 The brightest region in X-ray emission corresponds with the compact AGN source. The X-ray peak coincides with the highest SiO/CO and CN/CO intensity ratios measured in the CND (panels {\it c} and {\it d}). The contributions to the X-ray luminosity in the 0.25-7.50~keV band are known to be multiple (line and continuum emission, as well as emission from neutral and ionized material), and therefore the link between the molecular line ratios and the 'bolometric' X-ray flux, if any, is not easily interpretable. As argued below, by restricting the study to the hard X-ray band the interpretation is facilitated.
  
Figure.~\ref{SiO-CN-Xraystot}~(panels {\it b} and {\it c}) shows the superposition of the hard X-ray image obtained in the 6--8 keV band by Chandra (in contours) on the  SiO/CO and CN/CO ratio maps (color scale). The X-ray map has been obtained by applying an adaptive smoothing \footnote{using version 4.1 of CIAO package} on the Chandra archive NGC~1068 data available in this wave-band. This image is virtually identical to the hard X-ray map originally published and discussed by Ogle et al.~(\cite{Ogl03}; see also Young et al.~\cite{You01}). Emission in this band is dominated by the Fe I K$\alpha$ line, with a non negligible contribution from Fe XXV and scattered neutral and ionized continuum reflection (Ogle et al.~\cite{Ogl03}). Analysis of the 4--10~keV continuum observed with ASCA in NGC~1068 led Iwasawa et al.~(\cite{Iwa97}) to conclude that most of the hard X-ray flux comes from reflection by cold neutral gas. The strong emission detected in the  Fe I K$\alpha$ line in NGC~1068, a line generated by fluorescence in neutral cold molecular clouds, corroborates this picture.  Therefore, the 6-8~keV band image allows us to directly probe to what extent molecular gas in the CND of NGC~1068 is pervaded and processed by X-rays, without a significant contamination from other sources to the X-ray emission.

The hard X-ray emission in the inner 12$\arcsec \times$12$\arcsec$ FOV shown in Fig.~\ref{SiO-CN-Xraystot}~{\it b,c} is spatially resolved.  Young et al.~(\cite{You01}) confirmed the reality of this extended component, and concluded that the latter cannot be attributed  to scattering by the telescope PSF. The compact source, located on the AGN, is surrounded by lower-level emission that pervades the molecular material of the CND out to $r\sim$2--3$\arcsec$. This low-level emission shows a northeast protrusion beyond the CND extent, which aligns with the radio jet. The hard X-ray peak coincides within the errors with the highest SiO/CO and CN/CO intensity ratios measured in the CND, interpreted in Sect.~\ref{SiO-CN-CO-CND} in terms of a chemical enhancement (higher abundances) of SiO and CN. The picture of a correlation between X(SiO), X(CN) and the degree of X-ray irradiation of molecular gas is further illustrated in Fig.~\ref{Xrays-correlations}. There we have plot the SiO/CO and CN/CO ratios derived in  Sect.~\ref{SiO-CN-CO-CND} versus the X-ray flux measured, using  a 0.5$\arcsec$ grid over the CND and a common spatial resolution dictated in each case by the largest beam. Within an admittedly large scatter, Fig.~\ref{Xrays-correlations}  provides tentative evidence of a correlation. The correlation is tighter in the case of CN: the corresponding  coefficients of determination of the linear regressions are $r^2$=0.7 for CN and  $r^2$=0.6 for SiO.  
Based on the larger range of molecular line ratios derived when {\it high} and {\it low} velocities are separately analyzed (Sect.~\ref{SiO-CN-CO-CND}),  we speculate that the correlations of Fig.~\ref{Xrays-correlations} would have been found to be tighter if the spectral resolution of X-ray observations had allowed us to kinematically discriminate between the two components.

\subsubsection{X-ray chemistry in NGC~1068}\label{XDRchemistry}

The CND of NGC~1068 has the basic ingredients to become a giant XDR. First, the nucleus of this Seyfert 2 is a strong X-ray emitter. The intrinsic luminosity in the hard X-ray band has been estimated to be $\sim$10$^{43}$--10$^{44}$erg~s$^{-1}$ (e.g., Iwasawa et al.~\cite{Iwa97}; Maloney~\cite{Mal97}; Colbert et al.~\cite{Col02}). Furthermore, the central engine of NGC~1068 is surrounded by a massive ($\sim$5$\times$10$^{7}$M$_{\sun}$) circum-nuclear molecular disk of $\sim$200~pc radius. Thus, most of the molecular mass of the CND lies at a radius $\sim$100~pc where the expected hard X-ray flux is $\sim$10--100~erg~cm$^{-2}$~s$^{-1}$. This puts the molecular CND of NGC~1068 among the category of strongly irradiated XDRs (Meijerink et al.~\cite{Meij07}). As argued in Sect.~\ref{Xrays}, the hard X-ray image of NGC~1068 provides direct evidence that molecular gas in the CND is pervaded by X-rays. We therefore expect that the chemistry of molecular gas in the CND should show the footprints of X-ray processing predicted on theoretical grounds. As discussed below, there are specific predictions regarding SiO and CN.

Gas-phase models that analyze the chemistry of X-ray irradiated molecular gas foresee high abundances for CN, similar to those measured in the CND of NGC~1068: $X$(CN)$\sim$a few 10$^{-8}$--10$^{-7}$ (Lepp \& Dalgarno~\cite{Lep96}; Meijerink \& Spaans~\cite{Meij05}; Meijerink et al.~\cite{Meij07}). The abundance of CN is enhanced in XDR mainly as a result of the higher ionization degree of the gas. For similar reasons, there is theoretical basis supporting the enhancement of CN in PDR environments (e.g.; Boger \& Sternberg~\cite{Bog05}; Fuente et al.~\cite{Fue93, Fue08}; Janssen et al.~\cite{Jan95}). However, based on the age estimated for the most recent star formation episode taking place in the CND($>$2--3$\times$10$^8$~yrs old; Davies et al.~\cite{Dav09}),  and considering that the UV radiation from young massive stars should dominate the feedback in a starburst, we can safely discard  PDR chemistry as the origin for the high abundances of CN in the CND of NGC~1068. As argued in Sect.~\ref{30m-data}, shocks can also be rejected as an explanation for CN abundances.

The recent gas-phase models of  Meijerink et al.~(\cite{Meij07}) foresee a significant enhancement of SiO in strongly irradiated ($\sim$10--100~erg~cm$^{-2}$~s$^{-1}$) high density ($\geq$10$^{5}$cm$^{-3}$) XDRs, i.e., in a regime similar to the XDR environment of NGC~1068 where we measure $X$(SiO)$\sim$a few 10$^{-9}$--10$^{-8}$ . In particular,  Meijerink et al.~(\cite{Meij07}) predict that the SiO(1--0)/CO(1--0) ratio can reach a value of $\sim$0.10 assuming a standard depletion for silicon in gas phase. This is in close agreement with the average SiO(2--1)/CO(1--0) ratio observed in NGC~1068 ($\sim$0.08), assuming that the typical SiO(2--1)/SiO(1--0) ratios should be $\sim$1. The latter is an educated guess considering the physical conditions of SiO gas in the CND.

In addition to the predictions of gas-phase chemistry, Voit~(\cite{Voi91}) has proposed that the abundance of silicon monoxide could also be boosted due to dust grain processing by X-rays. These can evaporate small ($\sim$10~\AA) silicate grains and increase the Si fraction in gas phase, leading to a considerable enhancement of SiO in X-ray irradiated molecular gas (Mart\'{\i}n-Pintado et al.~\cite{Mar00};  U04; Garc{\'{\i}}a-Burillo et al.~\cite{Gar08}; Amo-Baladr\'on et al.~\cite{Amo09}). This scenario has been invoked by Mart\'{\i}n-Pintado et al.~(\cite{Mar00}) and Amo-Baladr\'on et al.~(\cite{Amo09}) to account for the correlation between the abundance of SiO and the equivalent width of the Fe K$\alpha$ fluorescence line in the Sgr A and Sgr B molecular cloud complexes of the Galactic Center. In the case of the Galactic Center, pure gas-phase XDR models have difficulties in reproducing the observed SiO abundances, unless an X-ray outburst is assumed to have taken place in Sgr~A$^*$ 300 yr ago. By contrast, current gas-phase XDR models successfully reproduce the SiO abundances derived in NGC~1068 without resorting to dust grain chemistry.  Nonetheless, the inclusion of dust grain chemistry, likely linked to the mechanical heating of the molecular ISM,  would help solve the controversy regarding the abundances of species like HCN and HNC, under-predicted by XDR schemes (Garc{\'{\i}}a-Burillo et al.~\cite{Gar08}; Loenen et al.~\cite{Loe08}; P\'erez-Beaupuits et al.~\cite{Per09}). Actually, while CN is overly produced in gas-phase XDR models, HCN is under-produced with respect to the level required by observations in AGNs. In addition to mechanical heating, the evaporation of dust grain mantles in dense hot environments like those likely prevailing in the nuclear disks of AGN, has been invoked as a mechanism responsible of enhancing HCN abundances (Lintott \& Viti~\cite{Lin06}).

 Other lines of evidence support the existence of an XDR in the CND. The low HCO$^{+}$/HOC$^{+}$  ratios measured by U04 can be explained only if molecular clouds have the high ionization degrees typical of XDR: $X$(e$^-$)$\sim$10$^{-6}$--10$^{-4}$. Furthermore, the excitation of the 2.12$\mu$m H$_2$ 
 rovibrational emission lines detected in the CND (Rotaciuc et al.~\cite{Rot91}; Blietz et al.~\cite{Blie94}; Galliano \& Alloin~\cite{Gal02}; M\"uller-S{\'a}nchez et al.~\cite{Mue09}) has been interpreted to be dominated by X-ray emission (e.g.; see discussion in Maloney~\cite{Mal97} and  Galliano \& Alloin~\cite{Gal02})

\section{Summary and Conclusions}\label{Summary and conclusions}

We have used the high spatial resolution and sensitivity capabilities of the PdBI to map the emission of the ($v$=0, $J$=2--1) line of SiO and the  $N$=2--1 transition of CN in the disk of the Seyfert 2 galaxy NGC~1068. The spatial resolution of these observations ($\sim$1--3$\arcsec$) have allowed us to separate the emission of the SB ring from that of the CND. The PdBI SiO and CN data, together with available PdBI CO maps of the galaxy, have been complemented by single-dish data obtained with the IRAM 30m telescope in SiO, CH$_ 3$OH and HNCO lines to probe the physical and chemical properties of molecular gas in the CND through an analysis of line ratios.

We summarize below the main results obtained in this work: 

\begin{itemize}

\item

Unlike the CO lines, which show strong emission in the SB ring, most of the SiO emission detected inside the PdBI primary beam comes from a circum-nuclear molecular disk (CND) located around the AGN. The dichotomy between the CND and the SB ring is reflected in the remarkably different $<R_\mathrm{SiO/CO} >$ ratios measured in the two regions. LVG models implemented to fit line ratios indicate that the average abundance of SiO in the CND, $<X$(SiO)$>$[CND]$\sim$(1--5)$\times$10$^{-9}$, is about one to two orders of magnitude higher than that measured in the SB ring. 

\item 

Similarly to CO, the SiO CND, of $\sim$400~pc deconvolved size, shows an asymmetric double peak structure oriented east-west.  The eastern knot corresponds to the SiO emission peak.  However, CO and SiO emissions show differences at the scales of the CND. The SiO/CO velocity-integrated intensity ratio changes by about a factor of three inside the CND, reaching a peak ratio of $\sim$0.10-0.12 towards the AGN and the western knot.

\item

The overall SiO kinematics in the CND are consistent with a rotating structure. However, there is evidence that the CND rotating pattern is distorted by non-circular and/or non-coplanar motions.  Compared to CO, SiO emission at small radii ($r\leq$2$\arcsec$=140~pc) stands out at  extreme velocities not accounted for by circular rotation.  The highest SiO/CO brightness temperature ratios in the CND are associated with the {\it high velocities} seen towards the AGN and the western knot. We interpret the spatial variations of the SiO/CO ratio in the CND as the likely signature of chemical differentiation. In this scenario, the abundance of SiO is significantly enhanced at {\it high velocities} out to $X$(SiO)$\sim$(0.3--1.5)$\times$10$^{-8}$.

\item

CN emission is also detected in a CND around the AGN. The size and morphology of the CND seen in CN are similar to those found in SiO. The overall abundance of CN in the CND is high: $<X$(CN)$>$[CND]$\sim$(0.2--1)$\times$10$^{-7}$. The CN/CO velocity-integrated intensity ratio changes by about a factor of three inside the CND and reaches a peak value of $\sim$0.30 towards the AGN. CN maps also show distorted kinematics that cannot be explained by circular rotation. The abundance of CN is significantly enhanced at {\it high velocities}, detected towards the AGN, out to  $X$(CN)$\sim$(0.8--4)$\times$10$^{-7}$.  The emission detected in CN at extreme velocities is physically associated with gas lying at small radii ($r\leq$1$\arcsec$=70~pc).
 
\item

Different models have been proposed to account for the kinematics of molecular gas in the CND. Large-scale shocks produced by cloud-cloud collisions in the ILR region of the NGC~1068 nuclear bar can explain the distorted kinematics and enhanced SiO abundances in the CND. The  hypothesis of a jet-ISM interaction, driving the large-scale expansion of the molecular ring, can also explain shocks. Independent evidence of shocks is provided by the detection of CH$_3$OH and HNCO. Line ratios involving these tracers and SiO in the CND are similar to those measured in prototypical shocked regions in our Galaxy. However, the strength and abundance of CN in NGC~1068 can be explained neither by shocks nor by PDR chemistry.  Alternatively, the high global abundances measured for  CN and SiO, in agreement with the theoretical predictions of XDR models, and the conspicuous correlation of CN/CO and SiO/CO ratios with the irradiation of hard X-rays, suggest that the CND of NGC~1068 has become a giant X-ray dominated region.

\end{itemize}

The extreme properties of molecular gas in the circum-nuclear disk of NGC~1068  result from the interplay between several processes directly linked to
nuclear activity.  Results presented in this paper highlight, in particular, the footprint of shocks and X-ray irradiation on the properties of molecular gas in this Seyfert. Whereas XDR chemistry offers a simple explanation for CN and SiO in NGC~1068, the relevance of shocks deserves further scrutiny, however. In particular, sub-arcsecond  spatial resolution observations, to be done at higher frequencies and in clear-cut molecular tracers, will be paramount to locate the occurrence of shocks in the CND and determine the gas mass and energies involved in the process. Increasing the spatial resolution at millimeter and sub-millimeter wavelengths will be also a key to explore in detail the correlation between molecular abundances and X-ray irradiation, tentatively identified in this work.  
Furthermore, increasing the number of molecular line transitions observed with high-spatial resolution will allow us to fit the different line ratios using multi-phase radiative transfer schemes. These are a better description of the multi-phase nature of the molecular ISM in galaxies. In particular, these models will properly take into account the fact that the emission of different lines can arise from regions characterized by very different densities and kinetic temperatures.

\begin{acknowledgements}
	 We acknowledge the IRAM staff from the Plateau de Bure and from 
         Grenoble for carrying out the observations and help provided during the 
	 data reduction. SGB and AF acknowledge support from MICIN within program CONSOLIDER INGENIO 2010, under grant 'Molecular Astrophysics: The Herschel and ALMA Era--
	 ASTROMOL' (ref CSD2009-00038).
      
\end{acknowledgements}

\end{document}